\documentclass[a4paper,fleqn]{cas-sc}

\usepackage[numbers,sort&compress]{natbib}
\usepackage{multicol}	
\usepackage{mathtools}  
\usepackage{amssymb} 
\usepackage{graphicx} 
\usepackage{placeins} 

\def\tsc#1{\csdef{#1}{\textsc{\lowercase{#1}}\xspace}}
\tsc{WGM}
\tsc{QE}

\begin{document}
\let\WriteBookmarks\relax
\def\floatpagepagefraction{1}
\def\textpagefraction{.001}

\shorttitle{STEM DP analysis with DL networks}    

\shortauthors{S. Wissel et~al.}  

\title [mode = title]{STEM Diffraction Pattern Analysis with Deep Learning Networks}  

\author[1]{Sebastian Wissel}[orcid=0000-0002-2179-1169]
\cormark[1] 
\ead{sebastian.wissel@tu-darmstadt.de}  
\credit{Investigation, Data curation, Methodology, Software, Visualization, Writing - original draft}
\author[2]{Jonas Scheunert}
\credit{Investigation}
\author[1]{Aaron Dextre} 
\credit{Investigation, Methodology, Software}
\author[2]{Shamail Ahmed} 
\credit{Investigation}
\author[2]{Andreas Beyer} 
\credit{Supervision}
\author[2]{Kerstin Volz} 
\cormark[1]
\ead{kerstin.volz@physik.uni-marburg.de}
\credit{Funding acquisition, Project administration, Supervision}
\author[1]{Bai-Xiang Xu} 
\cormark[1]
\ead{xu@mfm.tu-darmstadt.de}
\credit{Funding acquisition, Project administration, Supervision}

\affiliation[1]{organization={Technical University Darmstadt, Institute of Materials Science, Mechanics of Functional Materials},
            addressline={Otto-Berndt-Straße 3}, 
            city={Darmstadt},
            postcode={64287}, 
            state={Hessen},
            country={Germany}}
\affiliation[2]{organization={Philipps-University Marburg, Scientific Centre for Materials Science},
            addressline={Hans-Meerwein-Straße 6}, 
            city={Marburg},
            postcode={35032}, 
            state={Hessen},
            country={Germany}}

\cortext[1]{Corresponding authors.}

\begin{abstract}
Accurate grain orientation mapping is essential for understanding and optimizing the performance of polycrystalline materials, particularly in energy-related applications. Lithium nickel oxide (LiNiO\(_2\)) is a promising cathode material for next-generation lithium-ion batteries, and its electrochemical behaviour is closely linked to microstructural features such as grain size and crystallographic orientations. Traditional orientation mapping methods—such as manual indexing, template matching (TM), or Hough transform-based techniques—are often slow and noise-sensitive when handling complex or overlapping patterns, creating a bottleneck in large-scale microstructural analysis. This work presents a machine learning-based approach for predicting Euler angles directly from scanning transmission electron microscopy (STEM) diffraction patterns (DPs). This enables the automated generation of high-resolution crystal orientation maps, facilitating the analysis of internal microstructures at the nanoscale. Three deep learning architectures—convolutional neural networks (CNNs), Dense Convolutional Networks (DenseNets), and Shifted Windows (Swin) Transformers—are evaluated, using an experimentally acquired dataset labelled via a commercial TM algorithm. While the CNN model serves as a baseline, both DenseNets and Swin Transformers demonstrate superior performance, with the Swin Transformer achieving the highest evaluation scores and the most consistent microstructural predictions. The resulting crystal maps exhibit clear grain boundary delineation and coherent intra-grain orientation distributions, underscoring the potential of attention-based architectures for analyzing diffraction-based image data. These findings highlight the promise of combining advanced machine learning models with STEM data for robust, high-throughput microstructural characterization.
\end{abstract}

\begin{graphicalabstract}
\includegraphics[width=\textwidth]{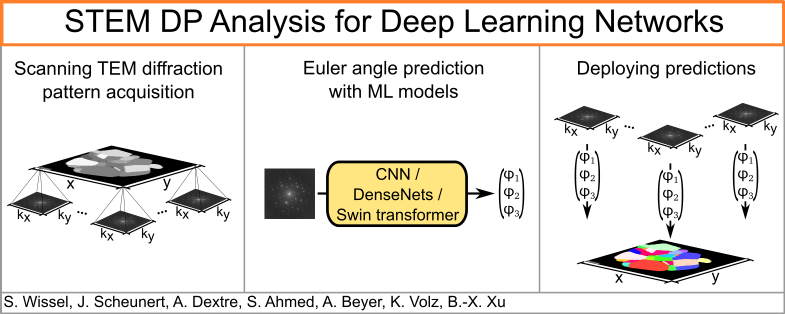}
\end{graphicalabstract}

\begin{highlights}
\item STEM diffraction pattern analysis
\item Grain orientation determination with ML models
\item Comparison of CNN, DenseNets and SwinT transformer
\item systematic hyperparameter study of the ML models
\end{highlights}

\maketitle

\section{Introduction}\label{s:intro}
Lithium nickel oxide LiNiO\(_2\) has emerged as a promising cathode material for next-generation lithium-ion batteries, offering high energy density and reduced reliance on cobalt. This shift aims to mitigate concerns related to cobalt’s limited availability, high cost, and ethical sourcing challenges~\cite{Manthiram2020,Schipper2016}. Understanding its microstructure is crucial, as grain orientations and grain boundaries influence electrochemical performance, mechanical stability, and degradation mechanisms~\cite{Xu2020,Bai2020,Chen2024,Chen2025}. Accurate characterization of these features is essential for optimizing battery performance and longevity.
Traditional methods for orientation mapping, such as manual indexing, TM or Hough transform-based techniques (primarily used for Kikuchi DPs), are often time-consuming and sensitive to noise, especially when dealing with complex or overlapping patterns~\cite{Williams2009,Rauch2014,Wright2005,Schwarzer2009}. These limitations pose a significant bottleneck in high-throughput or large-area microstructural analysis. 

Machine learning (ML) has emerged as a powerful tool for accelerating materials characterization by automating the analysis of complex microstructural data. In particular, deep learning models enable the rapid extraction of crystallographic information from imaging data, significantly reducing the time and effort required for conventional methods~\cite{DeCost2020,Jain2024,Kalinin2022,Ziletti2018,Butler2018}. The application of ML in microstructure investigations has gained increasing attention, particularly for energy materials, where efficient characterization is essential for materials development. As shown by \textit{Zuo et al.}~\cite{Zuo2022}, automated analysis of electron diffraction images using ML algorithms is fast and accurate, with the potential to substantially improve the efficiency and precision of materials characterization.
Building upon the general applicability of ML in crystallographic analysis, numerous studies have focused on applying CNNs to electron backscatter diffraction (EBSD) data from different materials~\cite{Jha2018,Suker2022,Ding2020,Hara2023,Shen2019,Yuan2021}. For instance, \textit{Jha et al.} demonstrated the effectiveness of CNNs in predicting grain orientations from EBSD data, achieving high accuracy and efficiency~\cite{Jha2018}. Similarly, \textit{Suker} applied deep learning techniques to EBSD patterns of AA5083 alloy, highlighting the potential of ML in microstructural characterization~\cite{Suker2022}. While these studies highlight the potential of ML-driven approaches for microstructural characterization, they predominantly rely on EBSD data, which is limited to Kikuchi DPs.

This study leverages STEM DPs, which enable nanoscale-resolution characterization of internal microstructures, providing deeper insights into crystallographic and microstructural features~\cite{Ding2020,Zuo2022}. Additionally, this study features the systematic evaluation of various ML architectures—including CNN, DenseNets, and Swin Transformers—for their effectiveness in analyzing STEM DPs, along with a systematic hyperparameter study of these models. CNN was selected as a widely adopted baseline in microstructural studies. DenseNets, which extend CNN by introducing dense connectivity and feature reuse, were included to explore potential gains in predictive accuracy. Swin Transformers offer a fundamentally different, attention-based architecture and were chosen for their recent success in outperforming CNNs on various computer vision tasks. This comprehensive assessment aims to explore and evaluate the strengths and weaknesses of the models for accurate grain orientation prediction and provides insights on model selection and hyperparameter optimization. 
By integrating STEM diffraction data with advanced ML models, this work demonstrates a pathway for high-resolution, automated orientation mapping and establishes a framework for future integration of artificial neural networks with electron microscopy in crystallographic analysis. The outcomes contribute to the broader goal of accelerating materials investigation and performance optimization, particularly in the context of energy-relevant materials.

\section{Methodology}\label{s:meth}
\subsection{Data preparation}\label{ss:data}
\begin{figure}
    \centering
    \includegraphics[width=\textwidth]{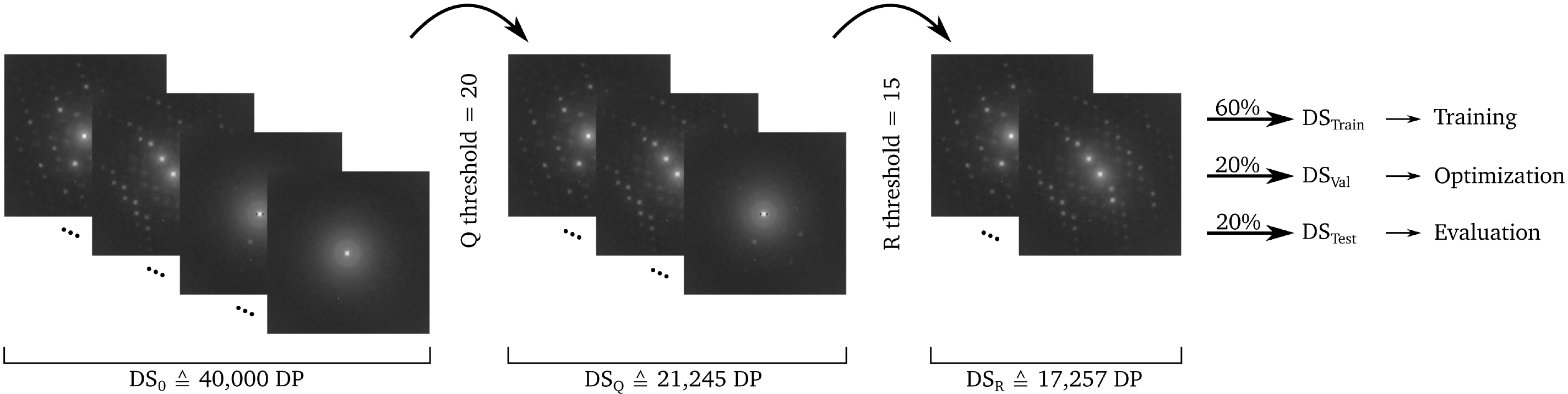}
    \caption{Training data ($DS_{Train}$) and evaluation data ($DS_{Val}$ and $DS_{Test}$) acquisition via DP and TM label classification.}
    \label{fig:DS_structure}
\end{figure}

The synthesis process of the monolithic LiNiO\(_2\) particles used in this study is mentioned in Ref.~\cite{Demuth2025}. An agglomerate of single-crystal LiNiO\(_2\) particles is selected for the sample preparation. The TEM lamella is prepared in a JEOL JIB 4601F focused ion beam microscope. First, a thin tungsten layer is deposited on the surface of the agglomerate using an electron beam followed by a thick tungsten layer with a Ga-ion beam to protect the surface of the particle during thinning. The agglomerate is attached to a micromanipulator needle and transferred to the TEM grid, where it is subsequently mounted. The lamella is thinned down using a Ga-ion beam. The final polishing is done using a 5 kV Ga-ion beam. The four-dimensional scanning precession electron diffraction (4D SPED) dataset is acquired using a JEOL 3010 transmission electron microscope operating at 300 kV, equipped with a NanoMegas P2010 system. The DPs are recorded using TemCam-XF416(ES) camera operated in 8× binning mode across the full sensor area. During dataset acquisition, a precession angle of $0.6^{\circ}$ and a semi-convergence angle of 1.6 mrad were used~\cite{Rauch2010}. From the acquired DPs, a virtual darkfield image was generated to provide an overview of the sampled region (see Fig.~\ref{fig:VDF_TMQ}~A).
After downsampling (image size from $512^2$ to $128^2$) and scaling the DP logarithmically, the orientation labels (Euler angles) are generated using the conventional cross-correlation-based TM method using ASTAR software version 2. For each DP, the correlation index $Q$ and orientation reliability $R$—which provide quantitative and qualitative assessments of the match—were also stored~\cite{Rauch2010,Rauch2014,Rauch2021}. DPs recorded outside the sample region typically yield significantly lower $Q$ values. To visualize label quality across the dataset, the TM labels weighted by their corresponding $Q$ values are shown in Fig.~\ref{fig:VDF_TMQ}~B.
Based on these quality metrics, the dataset was filtered by applying a $Q$ threshold of 20 to remove low-quality DPs and an $R$ threshold of 15 to discard DPs with unreliable labels. This filtering reduced the dataset from the original $DS_{0}$ (40,000 DPs and TM labels) to a refined dataset $DS_{R}$ (17,257 DPs and TM labels). The dataset $DS_{R}$ was then randomly split into training, validation, and test sets—$DS_{Train}$, $DS_{Val}$, and $DS_{Test}$—containing 60\%, 20\%, and 20\% of the data, respectively, for hyperparameter tuning and model evaluation (see Fig.~\ref{fig:DS_structure}).

As shown by \textit{Scheunert et al.}~\cite{Scheunert2025}, an issue arises by using the ASTAR software for labelling, which can not capture the intrinsic symmetry of the DPs. Since the templates are calculated by kinematic diffraction simulations, dynamic diffraction effects are neglected, especially effects in the zeroth-order Laue zone. As a result, certain crystallographic orientations--such as $\left[ 10\Bar{1}0 \right]$ and $\left[ 1\Bar{1}00 \right]$--produce near-identical templates and the TM algorithm can not distinguish between these orientations. This leads to systematic ambiguities in the assigned Euler angles, which affect the training of the ML models of this work. Furthermore, \textit{Scheunert et al.} showed that CNN algorithms are capable of recognising these intrinsic symmetries, which is also valid for the ML models of this work. Consequently, alternating predictions may occur within single grains that exhibit such symmetric orientations. This effect must be considered when interpreting the crystal maps presented in Section~\ref{s:res}.

The machine learning models of this work predict three Euler angles for a given DP, which describe the crystallographic orientation relative to a Cartesian coordinate system. These angles are transformed into grain orientations using the Python library $Orix$, which also facilitates the creation of inverse pole figures (IPFs)~\cite{Johnstone2020,Aanes2025}.

To fully describe the grain orientation, IPFs must be calculated with respect to all three coordinate axes (x, y, and z). Each axis-specific IPF provides a unique projection of the orientation. When visualized together, these IPFs represent a complete description of the orientation of the crystal.
In this work, the grain orientations are visualized primarily along the z-axis. For the sake of completeness, the remaining IPFs, calculated for the x- and y-axes, complement this analysis and can be found in the Appendix~\ref{a:IPFs}.

\begin{figure}[h]
    \centering
    \includegraphics[width=0.5\textwidth]{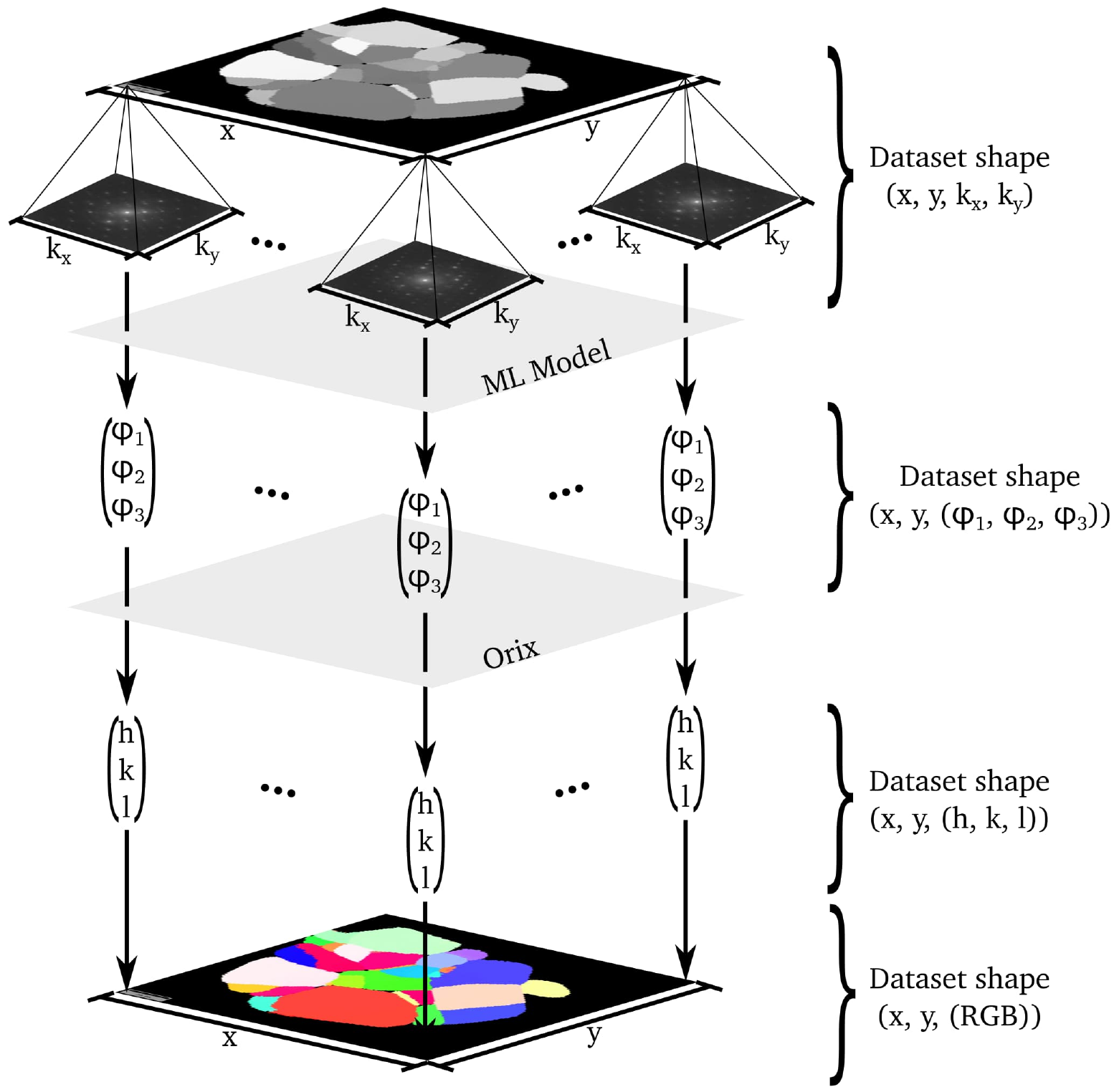}
    \caption{Workflow for the microstructure investigation.}
    \label{fig:microstructure}
\end{figure}

In addition to evaluation scores, microstructure comparison represents another key aspect of this study. As shown in Fig.~\ref{fig:microstructure}, the Euler angles of every DP of a whole dataset can be predicted, transformed into the corresponding grain orientation and visualized as a crystal map. Crystal maps are visualized using the materials' IPF colour code, allowing it to easily differentiate between different grains. Also, these crystal maps could be further analysed to extract microstructure parameters like grain size or texture.

\subsection{General ML training procedure}\label{ss:TrainingStructure}
\begin{figure}
    \centering
    \includegraphics[scale=1]{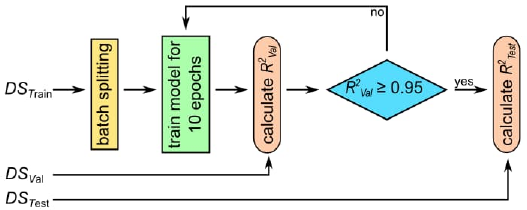}
    \caption{General training structure of trained ML models.}
    \label{fig:general_TS}
\end{figure}

The proposed machine learning models in this study follow a unified training structure (see Fig.~\ref{fig:general_TS}). The input data consists of individual, non-augmented DPs, while the targets are the three Euler angles obtained from the TM algorithm. The logarithmically scaled DPs are scaled linearly to the intensity range $[0, 1]$ for training. The target values are also normalized, to improve the gradient flow, activation functions performance and numerical stability.
The models are trained using the Huber loss function, which is robust to outliers by combining mean squared error (MSE) for small differences and mean absolute error (MAE) for larger deviations~\cite{Huber1964}. For learning rate optimization within an epoch, the Adam optimizer is used~\cite{Kingma2017}.
Model hyperparameters, such as batch size $BS$, learning rate $lr$, and - for convolutional neural networks - the number of convolutional layers and convolutional kernel size $CKS$, are optimized via a manual grid search. The number of training cycles, or epochs, is determined dynamically during training by using the two subsets $DS_{Train}$ and $DS_{Val}$. During training, $DS_{Train}$ is used to fit the model, while $DS_{Val}$ is predicted after every 10 epochs and the corresponding evaluation scores are calculated. Training stops when the $R^2$-score for $DS_{Val}$, $R^2_{Val}$, exceeds a predefined threshold, $R^2_{t} = 0.95$.
The final evaluation scores for each model are computed using predictions on the test dataset $DS_{Test}$, ensuring an unbiased assessment of model performance.
For the CNN and DenseNets models, a learning rate scheduler $LRS$ is tested. In this case, a constant $lr$ was replaced with an exponentially decreasing learning rate $lr_{i}$, where $i$ represents the current epoch and $lr_{0}$ the initial learning rate (see Eq.~\ref{eq:LRS}).

\begin{equation}
    lr_{i} = lr_{0} \cdot 0.995^{\frac{i+1}{500}}
    \label{eq:LRS}
\end{equation}

\subsection{Machine Learning models}\label{ss:MLmodels}
This section briefly overviews the ML models used, including the transfer learning strategy for DenseNets and SwinT transformers, and the employed CNN, DenseNet, and SwinT architectures.

Transfer learning is a machine learning technique that leverages pre-trained models to accelerate and improve the training process for a new task. Pre-trained models are first trained on a large and diverse dataset, capturing general patterns and features that can be adapted to specific problems with smaller, task-specific datasets. This approach is particularly beneficial when training from scratch is computationally expensive or when the available data for the new task is limited. In this study, we incorporated pre-trained models in the structures of some trained DenseNet and SwinT Transformer models. By using these pre-trained models as a foundation, we were able to build upon their learned feature representations, adapting them to our specific problem of predicting grain orientations from DPs. Additionally, this approach allowed us to compare the performance of ML models trained from scratch with those trained using pre-trained models. These comparisons provided insights into the effectiveness of different training strategies and enabled us to determine which model structure was best suited for our specific problem.

\subsubsection{Convolutional Neural Network}\label{sss:CNN}
\begin{figure}[h]
    \centering
    \includegraphics[width=0.5\textwidth]{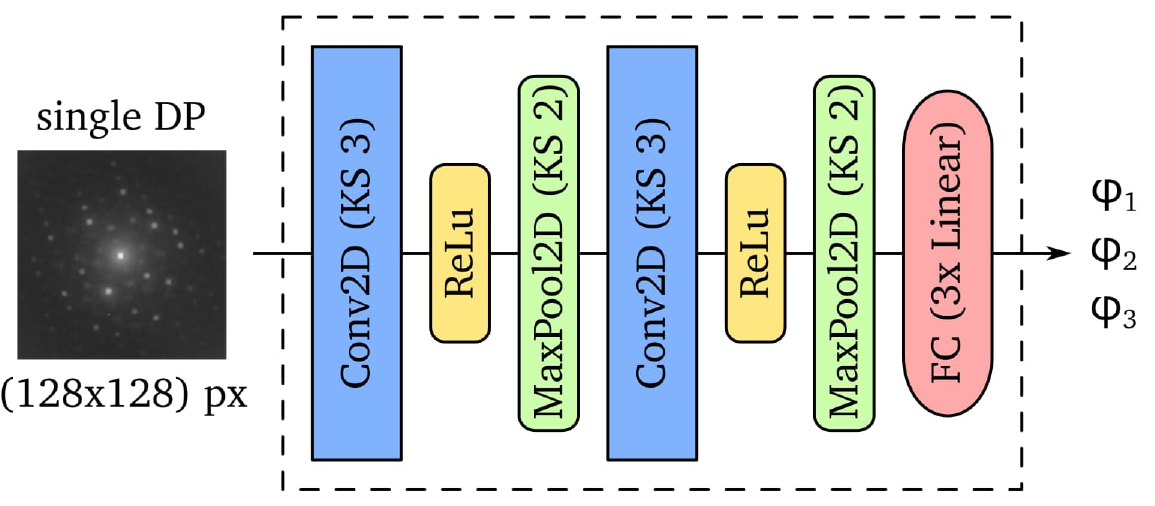}
    \caption{Structure of trained CNN models.}
    \label{fig:CNN_TS}
\end{figure}

Convolutional Neural Networks are a type of deep learning model designed to process data with a grid-like topology, such as images. They utilize convolutional layers to extract spatial features from input data by applying learnable filters (kernels). These filters slide over the input, detecting patterns like edges or textures, which are then passed through activation functions and pooling layers to reduce dimensionality while retaining the most relevant features. CNNs are particularly effective for image-based tasks due to their ability to learn hierarchical representations.~\cite{Indolia2018,Oshea2015,Li2022}\\
The CNN trained in this work consists of two convolutional layers, each followed by a ReLU activation function and a 2D max pooling operation. After feature extraction, the output is passed through three fully connected (FC) layers, which map the features to the target output: an array of the three Euler angles. The structure of the trained CNN is illustrated in Fig.~\ref{fig:CNN_TS}. 

\subsubsection{Dense Convolutional Network}\label{sss:DNs}
\begin{figure}[h]
    \centering
    \includegraphics[width=\textwidth]{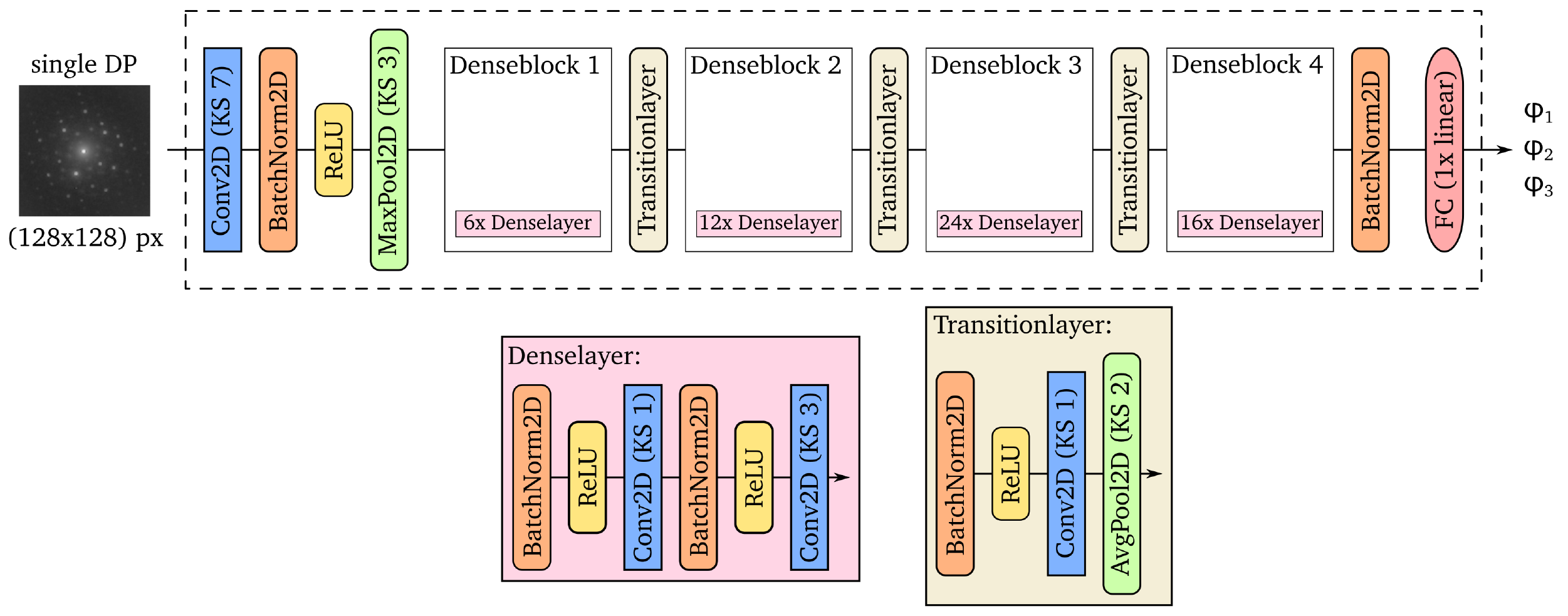}
    \caption{Structure of trained DenseNets models, including Denselayer and Transitionlayer structure.}
    \label{fig:Densenets_TS}
\end{figure}

Dense Convolutional Networks are an extension of CNNs, designed to improve feature propagation and reduce the number of parameters. In DenseNets, each layer is connected to every subsequent layer within the same block. This structure ensures that features learned by earlier layers are reused in later layers, which promotes efficient feature learning and mitigates the vanishing gradient problem. Compared to standard CNNs, DenseNets offer improved computational efficiency and performance, especially for image-based tasks.~\cite{Huang2018}\\
The DenseNets models trained in this work follow the DenseNet-121 architecture. Two variants were employed: a pre-trained model and a model trained from scratch (plain model). The models share the same structure (Fig.~\ref{fig:Densenets_TS}), which starts with an initial convolution, followed by four dense blocks with 6, 12, 24 and 16 dense layers, respectively. In between the dense blocks are three transition layers and the model structure ends with a batch normalization and a single FC layer. In both cases, the input is a single DP, and the output is an array of three Euler angles.

\subsubsection{Shifted windows Transformer}\label{sss:SwinT}
\begin{figure}[h]
    \centering
    \includegraphics[width=\textwidth]{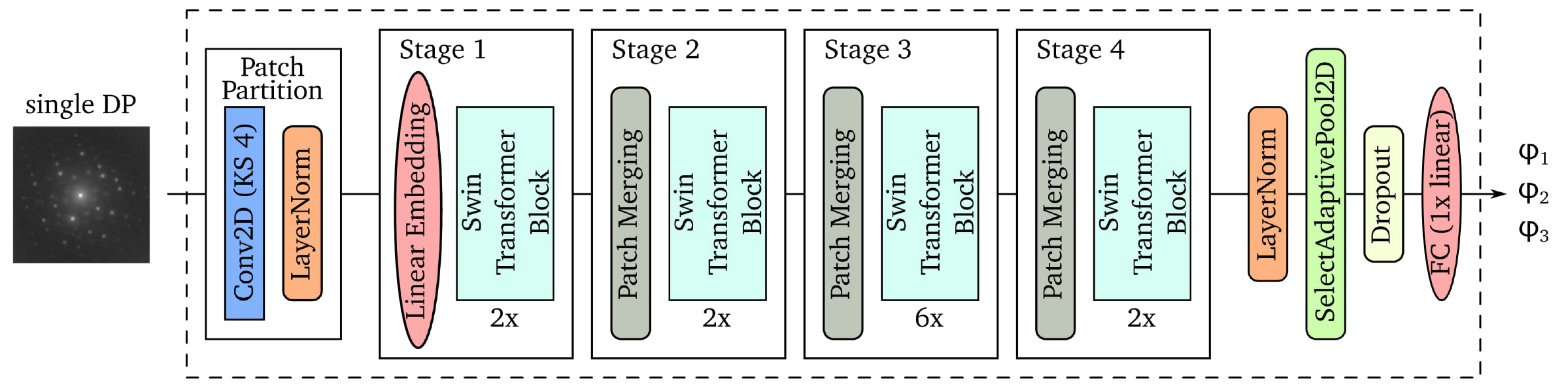}
    \caption{Structure of trained SwinT transformer models.}
    \label{fig:SwinT_TS}
\end{figure}

Unlike CNNs and DenseNets, which rely on convolutional filters, Shifted windows Transformers utilize self-attention mechanisms to capture long-range dependencies in the input data. The Swin Transformer is a deep learning architecture designed to handle image-based tasks. By dividing the input image into non-overlapping patches and applying attention within local windows, the Swin Transformer achieves computational efficiency while maintaining a global understanding of the input. Compared to CNNs, Swin Transformers are particularly advantageous for tasks requiring high-level feature representations and global context.~\cite{Liu2021,Liu2022}\\
In this work, we trained four Swin Transformer models of version \textit{SwinT}. The two pre-trained models share the same structure (see Fig.~\ref{fig:SwinT_TS}), with only the input image size differing. In the version "adjust images", the size of the DPs is increased to match the model input size of $224^2$. In the version "adjust model", the model input size is set to the DP size of $128^2$. The first plain SwinT model, called "modified FC layer", has the same structure as the pre-trained models, where the FC layer consists of a single linear layer. The second plain SwinT model, called "custom FC layer", has one variance in the structure, as its FC layer consists of two linear layers. In all cases the output is an array of three Euler angles.

\FloatBarrier

\section{Results and Discussion}\label{s:res}
\begin{figure}
    \centering
    \includegraphics[width=0.75\textwidth]{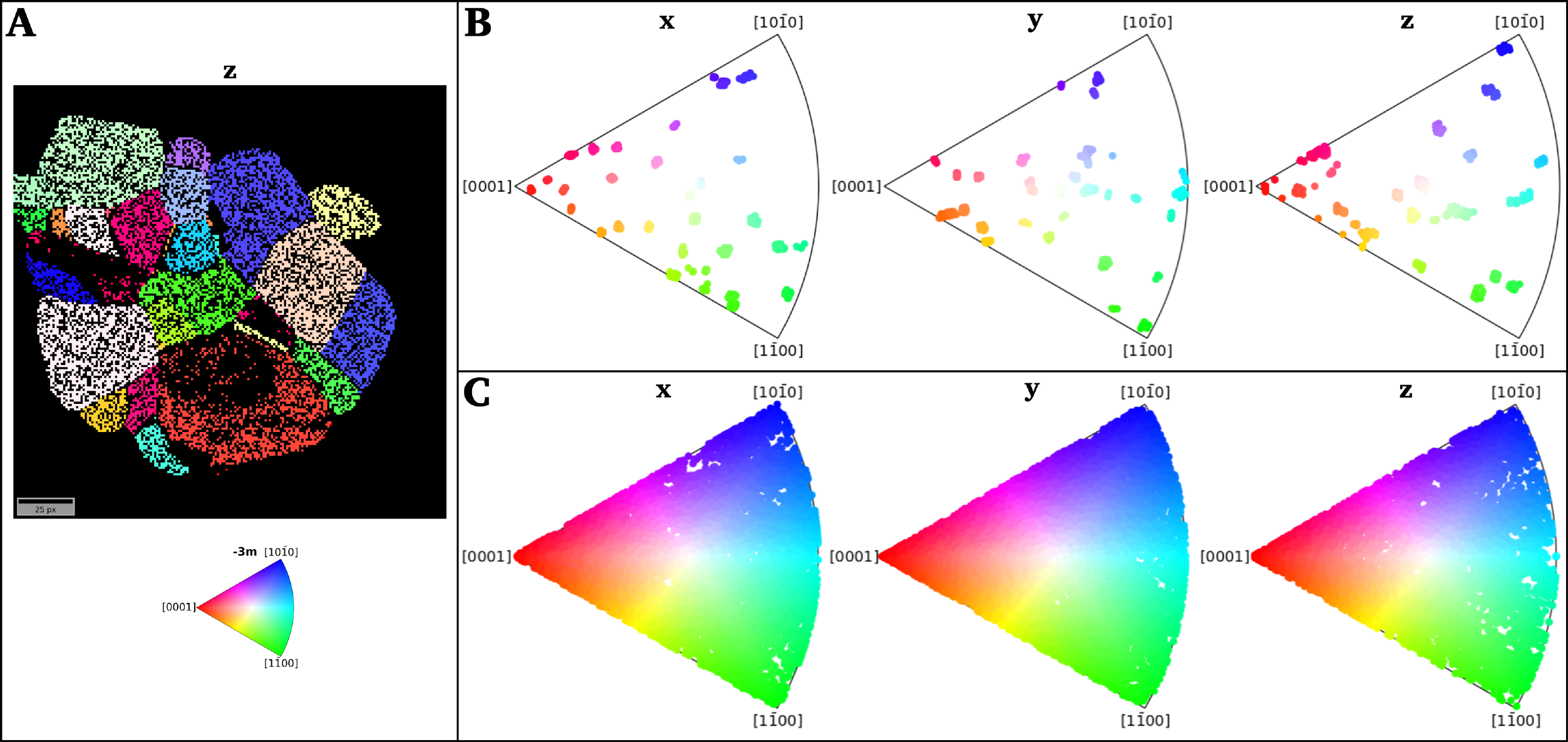}
    \caption{The TM labels of $DS_{Train}$ shown as crystal map (A) and scatterplot (B) and the predictions of a trained CNN for $DS_{Q}$ as scatterplot (C).}
    \label{fig:TD_CNN_pred_scatter}
\end{figure}

To assess the representativeness of the training data and evaluate the model's ability to generalize beyond the specific orientations seen during training, the distribution of grain orientations in the training dataset $DS_{Train}$ was first analyzed. As shown in Fig.~\ref{fig:TD_CNN_pred_scatter}~A and B, the training dataset contains a diverse, but not uniform, distribution of orientations. This is expected, as the dataset is based on experimental measurements and therefore does not cover the entire orientation space.

To verify whether this partial sampling of orientation space limits the predictive capabilities of the trained model, the entire dataset $DS_{Q}$ was predicted using a trained CNN model. The resulting predicted orientations are visualized as scatter plots in Fig.~\ref{fig:TD_CNN_pred_scatter}~C. These plots demonstrate that the model is capable of generating a wide range of grain orientations, extending beyond those explicitly represented in the training data. This indicates that the CNN model successfully learned a generalized mapping between DPs and crystal orientations, rather than memorizing specific training samples.

\begin{figure}
    \centering
    \includegraphics[width=0.5\textwidth]{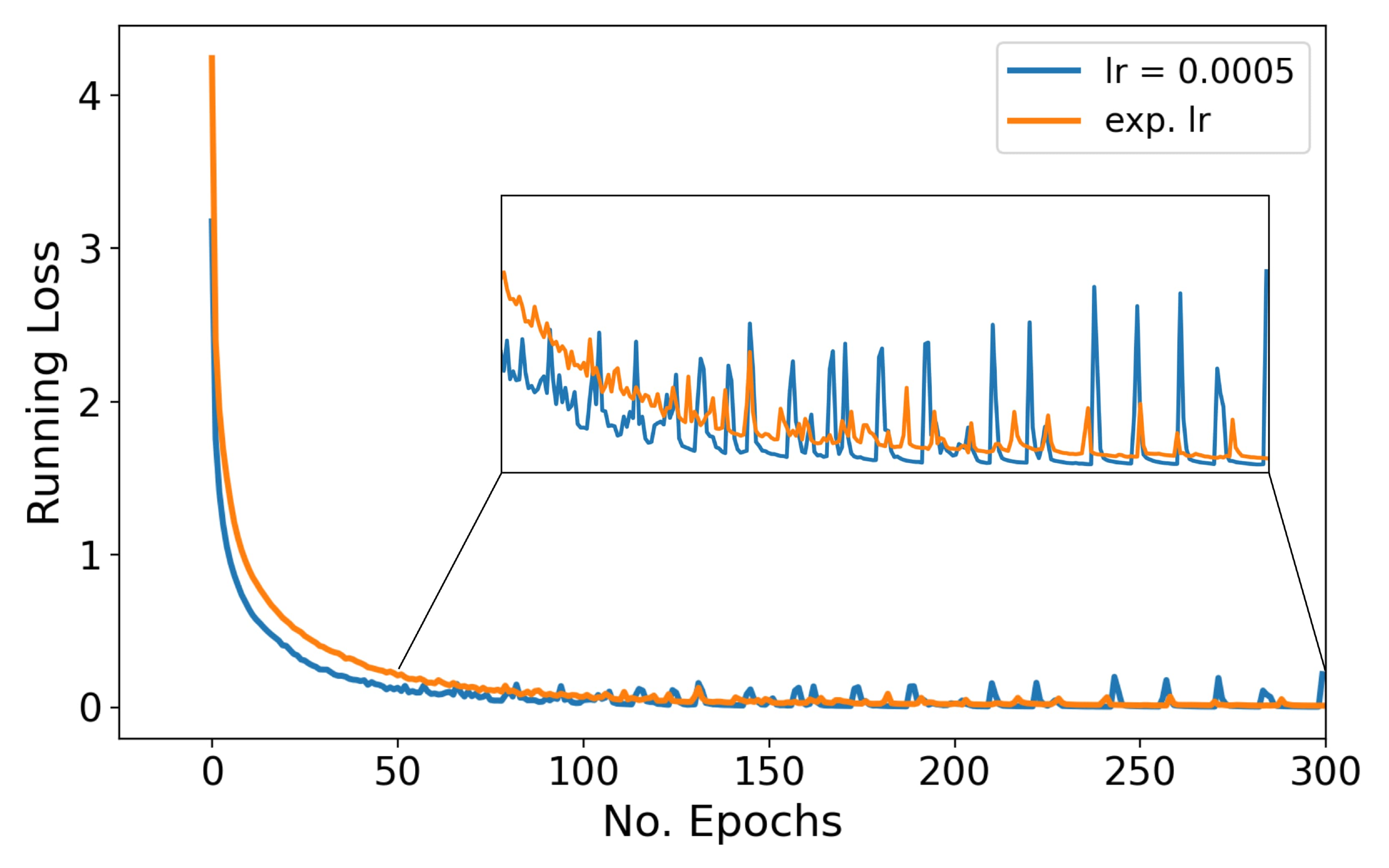}
    \caption{Training loss development for a CNN model trained with constant $lr = 0.0005$  and exponentially decreasing $lr$.}
    \label{fig:comp_TL}
\end{figure}

For performance optimization of the CNN model, an extensive hyperparameter tuning process was conducted. Although optimization mainly focused on the parameters batch size ($BS$), convolutional kernel size ($CKS$) and learning rate scheduling ($LRS$), initially the impact of the number of convolutional layers was investigated. Starting with four convolutional layers, the network accuracy was significantly improved by the reduction to two convolutional layers.
The influence of $BS$ was analyzed while keeping a constant learning rate ($lr = 0.005$). Models trained with small batch sizes ($BS = 8, 16, 32$) exhibited short training times but failed to achieve satisfactory performance. Conversely, models trained with larger batch sizes ($BS = 96, 128, 256$) resulted in significantly longer training times without a notable improvement in model accuracy. Notably, the model trained with $BS = 256$ performed significantly worse than the others. Based on these observations, an intermediate batch size of $BS = 64$ was selected, as it provided the best trade-off between training time and model performance, requiring fewer epochs to reach optimal performance.
In contrast, clamping the output to a physically reasonable range before calculating the loss led to significantly worse network performance and was excluded in subsequent hyperparameter studies.
In addition, the impact of the $CKS$ was examined (constant $lr = 0.005$). \textit{Ding et al.} recommend the use of large convolutional kernels for improved feature extraction and model performance~\cite{Ding2022}. To evaluate if this holds for the present task, CNN models were trained with different kernel sizes ($CKS = 3, 7, 13, 19, 25, 31$), while all other hyperparameters were kept constant. All trained models achieved similar evaluation scores, but larger kernels resulted in significantly longer training times. Due to missing performance improvement and given their computational cost, the conventional kernel size of $3$ was chosen, rather than a larger $CKS$.
While training with a constant $lr$, multiple models showed loss fluctuations at higher epoch numbers. To stabilize the training process, learning rate scheduling was investigated and the constant $lr$ was replaced with an exponentially decreasing $lr_{i}$ (Eq.~\ref{eq:LRS}).As an initial learning rate $lr_{0} = 0.0005$ was chosen. This adjustment resulted in a more stable training process and improved convergence (Fig.~\ref{fig:comp_TL}), which also enhanced the model’s performance (Tab.~\ref{tab:eval_scores_lrs}).
The CNN model trained with a $CKS$ of 3, $BS$ of 64 and the $LRS$ complied with the quality criterion after 100 epochs of training.
For performance optimization of the DenseNets models, the impact of $BS$ and $lr$ was evaluated. For both the plain and pre-trained DenseNets architectures, a batch size of $BS = 64$ combined with the $LRS$ using an initial learning rate of $lr_{0} = 0.0005$ yielded the best results. These configurations met the defined quality criterion after 70 epochs for the plain model and 60 epochs for the pre-trained variant.
In contrast, the more complex SwinT transformer models were trained with a constant $lr$ of 0.0001 and a $BS$ of 32. Training duration varied significantly among the four SwinT variants. The "adjust images", "adjust model", and "modified FC layer" versions fulfilled the stopping criterion within 70, 60, and 80 epochs, respectively. However, the "custom FC layer" model required 510 epochs to converge.

\begin{table}
\centering
\caption{Evaluation scores of CNN models trained with constant and exponentially decreasing $lr$}
\label{tab:eval_scores_lrs}
\setlength{\tabcolsep}{4pt} 
\renewcommand{\arraystretch}{1.1} 
\begin{tabular}{lccc}
\toprule
 $lr$ & $MSE$ & $MAE$ & $R_{DS_{Q}}^{2}$ \\ 
\midrule
$0.0005$ (const.) & 359.73 & 9.46 & 0.94 \\
exp. $lr$ & 219.41 & 5.55 & 0.97 \\
\bottomrule
\end{tabular}
\end{table}

The performance of the three best-performing models is shown in Table~\ref{tab:eval_scores_MLmodels}. The DenseNets and SwinT models achieve slightly higher $R^2$-scores than the CNN when predicting $DS_{Test}$ of the training-validation-test split. However, all three models perform equally well when predicting the entire $DS_{Q}$ dataset. The lower $R^2$ scores can be attributed to the prediction of DP with unreliable TM labels, which decreases the meaningfulness of the evaluation scores.

\begin{figure}
    \centering
    \includegraphics[width=0.9\textwidth]{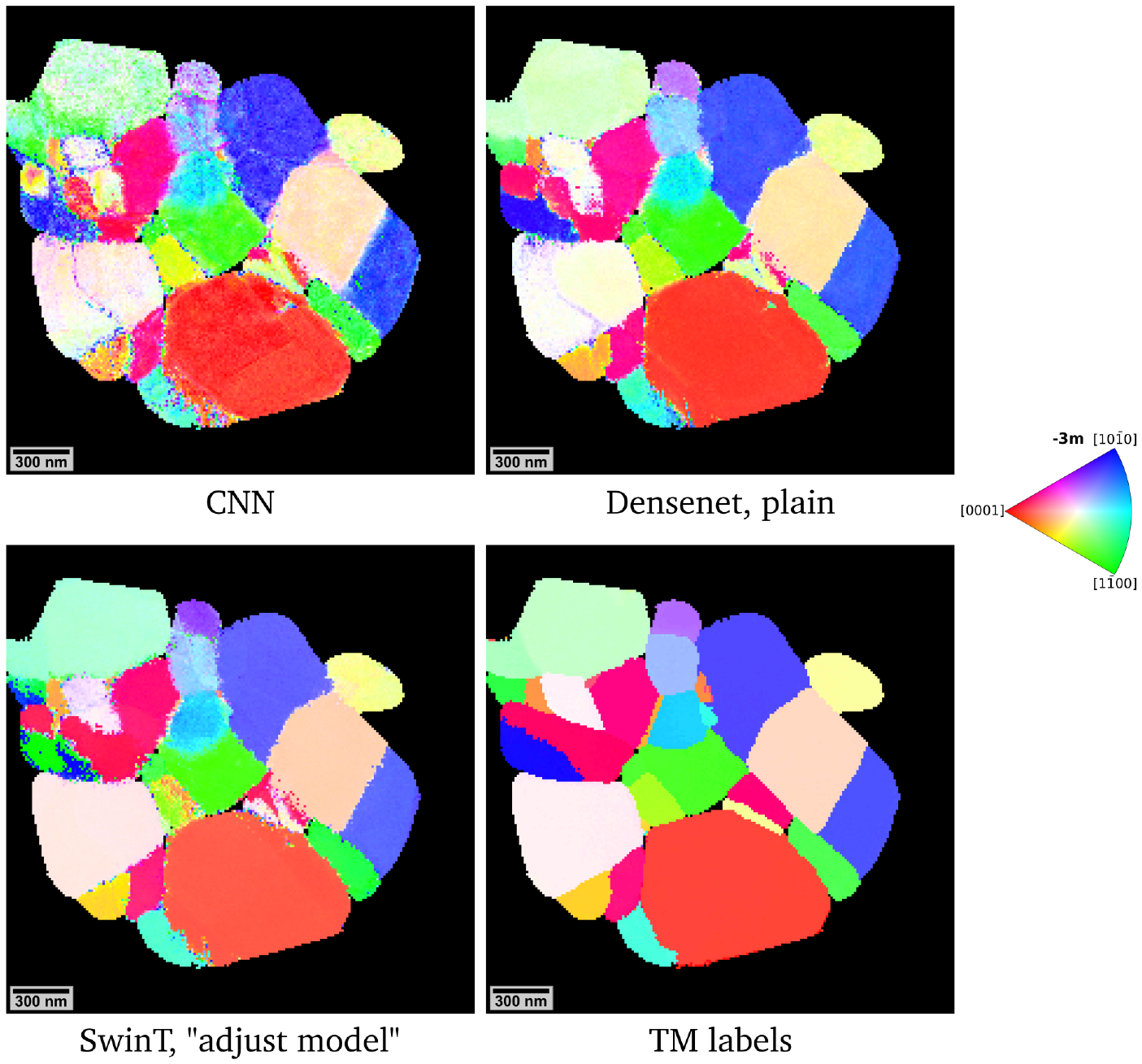}
    \caption{Visual comparison of the prediction of $DS_{Q}$ with the best trained ML models and the TM labels.}
    \label{fig:comp_MLmodels}
\end{figure}

Besides the performance assessment by the evaluation scores, which is a plain mathematical evaluation, the analysis of the microstructure prediction assesses the model performance concerning physically meaningfulness. As shown in Fig.~\ref{fig:microstructure}, transforming the predicted Euler angles of an entire dataset into the corresponding grain orientations and visualizing these in the IPF colour code, leads to crystal maps. In Fig.~\ref{fig:comp_MLmodels}, the visual comparison of the model predictions and the TM labels is shown. Overall, the predictions are similar and assign similar grain orientations to the same regions. They vary in terms of grain boundary sharpness and prediction consistency within individual grains. While the CNN prediction appears granular and inconsistent, the DenseNets prediction shows sharper grain boundaries and less divergent predictions within grains. Both are outperformed by the SwinT transformer, which shows the best performance in both categories. The generated crystal maps also allow for the extraction of various microstructural parameters, such as grain size distribution, volume fraction, texture, and grain boundary characteristics, enabling a comprehensive crystallographic analysis of the sample.

\begin{table}
\centering
\caption{Hyperparameters of ML models and corresponding evaluation scores for different Datasets}
\label{tab:eval_scores_MLmodels}
\setlength{\tabcolsep}{4pt} 
\renewcommand{\arraystretch}{1.1} 
\begin{tabular}{l|ccc|ccc}
\toprule
 ML model & $BS$ & $lr_0$ & $LRS$ & $R_{Val}^{2}$ & $R_{Test}^{2}$ & $R_{DS_{Q}}^{2}$ \\ 
\midrule
CNN & 64 & 0.0005 & \textit{yes} & 0.95 & 0.94 & 0.92 \\
Densenet, plain & 64 & 0.0005 & \textit{yes} & 0.96 & 0.95 & 0.92 \\
SwinT, pre-trained adjusted model & 32 & 0.0001 & \textit{no} & 0.96 & 0.95 & 0.92 \\
\bottomrule
\end{tabular}
\end{table}

\begin{figure}
    \centering
    \includegraphics[width=\textwidth]{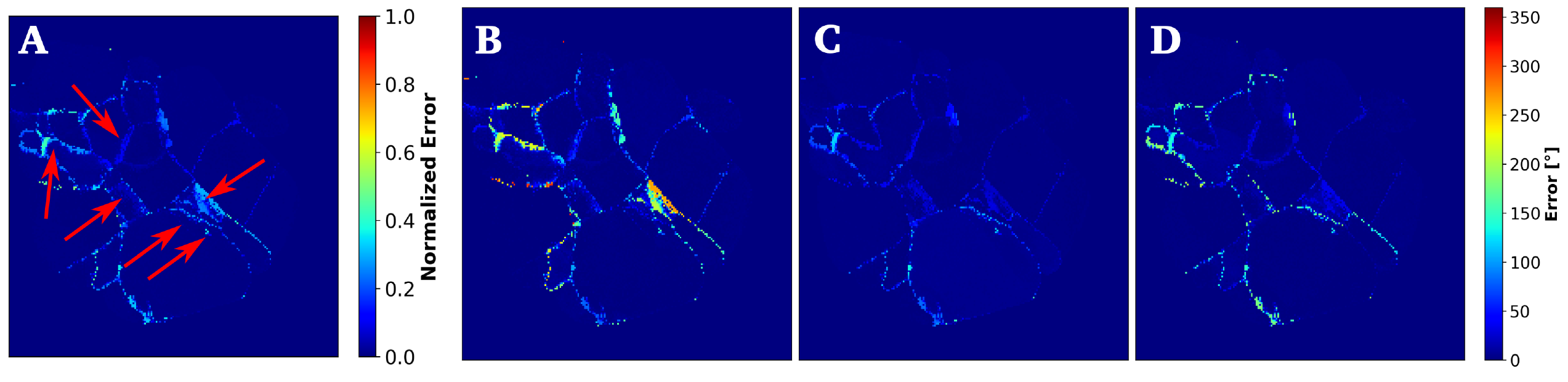}
    \caption{Normalized MAE with highlighted deviations (A) and absolute error of the three predicted Euler angles (B-D) of the best trained SwinT model compared with TM labels.}
    \label{fig:error_SwinT}
\end{figure}

The high similarity between the SwinT models' microstructure prediction and the ground truth TM labels (see Fig.~\ref{fig:comp_MLmodels}) demonstrates the precise prediction capability of the ML model. Several minor differences between the predictions and TM labels can be attributed to the limitations of the TM labels, based on their kinematic simulation (see Sec.~\ref{ss:data}).
The error distribution, shown in Fig.~\ref{fig:error_SwinT}, reveals that the largest deviations occur in smaller grains and at grain boundaries. While smaller grains are less represented in the training dataset, the TM labels in these regions also exhibit lower $Q$ and $R$ scores. This indicates that the ground truth itself is of lower quality, making variations between the predictions and TM labels less meaningful for assessing the ML model’s accuracy. Moreover, due to the lower reliability of TM labels at grain boundaries, the ML model predictions may, in some cases, be closer to the true grain orientations than the TM results.
Furthermore, the ML model has another advantage over the TM method, as it is sensitive to overlapping grains. As highlighted in Fig.~\ref{fig:error_SwinT}~A, some deviations between TM and predictions appear inside grains. Since the TM method assigns labels based on the dominant crystal orientation, there is an abrupt alteration if the dominant crystal orientation changes. However, the SwinT model is sensitive to the smooth transition of dominant crystal orientation and assigns slightly adjusted labels.

\FloatBarrier

\section{Conclusion}\label{s:con}
Using the proposed pipeline, this study demonstrates that machine learning models can accurately predict Euler angles from STEM DPs. These predictions enable the automated generation of crystal orientation maps, which provide detailed insight into the microstructure of the analysed samples. While all tested models achieved strong predictive performance, the CNN served as a baseline. Both advanced architectures—DenseNets and SwinT transformers—outperformed the CNN, with the SwinT model featuring an adjusted architecture showing the best overall performance. It achieved the highest evaluation scores and generated crystal maps with sharp grain boundaries and consistent intra-grain predictions. Furthermore, it showed sensitivity to overlapping grains, which is an advantage compared to the previously used TM method. These results suggest that attention-based architectures are particularly well-suited for analyzing DPs and similarly complex image data due to their ability to capture long-range spatial dependencies.

In addition to architectural comparisons, a detailed hyperparameter study-conducted primarily on the CNN—revealed further insights. The results highlight that training with high-quality labels, obtained by sorting the dataset according to the confidence scores of the TM method, significantly enhances predictive accuracy. Furthermore, both batch size and the choice of learning rate schedule were found to play a crucial role in training stability and overall performance. An intermediate batch size combined with an exponentially decreasing learning rate resulted in faster convergence and more reliable predictions. Conversely, applying output clamping before the loss calculation and increasing the convolutional kernel size did not improve performance and were therefore not included in the final setup.

The presented approach has potential to accelerate and enhance microstructure characterization in materials science. Future work will investigate the influence of sample thickness and improve model training by leveraging simulated DPs with highly accurate orientation labels. This will help eliminate potential indexing errors introduced by the TM method and further refine the model accuracy.
\bigskip

\printcredits

\section*{Declaration of competing interest}
The authors declare that they have no known competing financial interests or personal relationships that could have appeared to influence the work reported in this paper.

\section*{Acknowledgements}
The authors gratefully acknowledge the computing time provided to them on the high-performance computer Lichtenberg II at TU Darmstadt, funded by the German Federal Ministry of Education and Research (BMBF) and the State of Hesse.

\newpage
\bibliographystyle{cas-model2-names}


\begin{thebibliography}{37}
\expandafter\ifx\csname natexlab\endcsname\relax\def\natexlab#1{#1}\fi
\providecommand{\url}[1]{\texttt{#1}}
\providecommand{\href}[2]{#2}
\providecommand{\path}[1]{#1}
\providecommand{\DOIprefix}{doi:}
\providecommand{\ArXivprefix}{arXiv:}
\providecommand{\URLprefix}{URL: }
\providecommand{\Pubmedprefix}{pmid:}
\providecommand{\doi}[1]{\href{http://dx.doi.org/#1}{\path{#1}}}
\providecommand{\Pubmed}[1]{\href{pmid:#1}{\path{#1}}}
\providecommand{\bibinfo}[2]{#2}
\ifx\xfnm\relax \def\xfnm[#1]{\unskip,\space#1}\fi
\bibitem[{Manthiram(2020)}]{Manthiram2020}
\bibinfo{author}{Manthiram, A.}, \bibinfo{year}{2020}.
\newblock \bibinfo{title}{A reflection on lithium-ion battery cathode
  chemistry}.
\newblock \bibinfo{journal}{Nature Communications} \bibinfo{volume}{11},
  \bibinfo{pages}{1550}.
\newblock \URLprefix \url{https://doi.org/10.1038/s41467-020-15355-0},
  \DOIprefix\doi{10.1038/s41467-020-15355-0}.
\bibitem[{Schipper et~al.(2016)Schipper, Erickson, Erk, Shin, Chesneau and
  Aurbach}]{Schipper2016}
\bibinfo{author}{Schipper, F.}, \bibinfo{author}{Erickson, E.M.},
  \bibinfo{author}{Erk, C.}, \bibinfo{author}{Shin, J.Y.},
  \bibinfo{author}{Chesneau, F.F.}, \bibinfo{author}{Aurbach, D.},
  \bibinfo{year}{2016}.
\newblock \bibinfo{title}{Review—recent advances and remaining challenges for
  lithium ion battery cathodes}.
\newblock \bibinfo{journal}{Journal of The Electrochemical Society}
  \bibinfo{volume}{164}, \bibinfo{pages}{A6220}.
\newblock \URLprefix \url{https://dx.doi.org/10.1149/2.0351701jes},
  \DOIprefix\doi{10.1149/2.0351701jes}.
\bibitem[{Xu et~al.(2020)Xu, Jiang, Kuai, Xu, Qin, Zhang, Rahman, Wei,
  Nordlund, Sun, Xiao, Du, Zhao, Yan, Liu and Lin}]{Xu2020}
\bibinfo{author}{Xu, Z.}, \bibinfo{author}{Jiang, Z.}, \bibinfo{author}{Kuai,
  C.}, \bibinfo{author}{Xu, R.}, \bibinfo{author}{Qin, C.},
  \bibinfo{author}{Zhang, Y.}, \bibinfo{author}{Rahman, M.M.},
  \bibinfo{author}{Wei, C.}, \bibinfo{author}{Nordlund, D.},
  \bibinfo{author}{Sun, C.J.}, \bibinfo{author}{Xiao, X.}, \bibinfo{author}{Du,
  X.W.}, \bibinfo{author}{Zhao, K.}, \bibinfo{author}{Yan, P.},
  \bibinfo{author}{Liu, Y.}, \bibinfo{author}{Lin, F.}, \bibinfo{year}{2020}.
\newblock \bibinfo{title}{Charge distribution guided by grain crystallographic
  orientations in polycrystalline battery materials}.
\newblock \bibinfo{journal}{Nature Communications} \bibinfo{volume}{11},
  \bibinfo{pages}{83}.
\newblock \URLprefix \url{https://doi.org/10.1038/s41467-019-13884-x},
  \DOIprefix\doi{10.1038/s41467-019-13884-x}.
\bibitem[{Bai et~al.(2020)Bai, Zhao, Liu, Stein and Xu}]{Bai2020}
\bibinfo{author}{Bai, Y.}, \bibinfo{author}{Zhao, K.}, \bibinfo{author}{Liu,
  Y.}, \bibinfo{author}{Stein, P.}, \bibinfo{author}{Xu, B.X.},
  \bibinfo{year}{2020}.
\newblock \bibinfo{title}{A chemo-mechanical grain boundary model and its
  application to understand the damage of li-ion battery materials}.
\newblock \bibinfo{journal}{Scripta Materialia} \bibinfo{volume}{183},
  \bibinfo{pages}{45--49}.
\newblock \URLprefix
  \url{https://www.sciencedirect.com/science/article/pii/S1359646220301597},
  \DOIprefix\doi{https://doi.org/10.1016/j.scriptamat.2020.03.027}.
\bibitem[{Chen et~al.(2024)Chen, Allen, Rezaei, Furat, Schmidt, Singh, Weddle,
  Smith and Xu}]{Chen2024}
\bibinfo{author}{Chen, W.X.}, \bibinfo{author}{Allen, J.M.},
  \bibinfo{author}{Rezaei, S.}, \bibinfo{author}{Furat, O.},
  \bibinfo{author}{Schmidt, V.}, \bibinfo{author}{Singh, A.},
  \bibinfo{author}{Weddle, P.J.}, \bibinfo{author}{Smith, K.},
  \bibinfo{author}{Xu, B.X.}, \bibinfo{year}{2024}.
\newblock \bibinfo{title}{Cohesive phase-field chemo-mechanical simulations of
  inter- and trans- granular fractures in polycrystalline nmc cathodes via
  image-based 3d reconstruction}.
\newblock \bibinfo{journal}{Journal of Power Sources} \bibinfo{volume}{596},
  \bibinfo{pages}{234054}.
\newblock \URLprefix
  \url{https://www.sciencedirect.com/science/article/pii/S0378775324000053},
  \DOIprefix\doi{https://doi.org/10.1016/j.jpowsour.2024.234054}.
\bibitem[{Chen et~al.(2025)Chen, Peng, Wu, Furat, Schmidt and Xu}]{Chen2025}
\bibinfo{author}{Chen, W.X.}, \bibinfo{author}{Peng, X.L.},
  \bibinfo{author}{Wu, J.Y.}, \bibinfo{author}{Furat, O.},
  \bibinfo{author}{Schmidt, V.}, \bibinfo{author}{Xu, B.X.},
  \bibinfo{year}{2025}.
\newblock \bibinfo{title}{A length-scale insensitive cohesive phase-field
  interface model: Application to concurrent bulk and interface fracture
  simulation in lithium-ion battery materials}.
\newblock \bibinfo{journal}{Journal of the Mechanics and Physics of Solids}
  \bibinfo{volume}{196}, \bibinfo{pages}{106013}.
\newblock \URLprefix
  \url{https://www.sciencedirect.com/science/article/pii/S0022509624004794},
  \DOIprefix\doi{https://doi.org/10.1016/j.jmps.2024.106013}.
\bibitem[{Williams and Carter(2009)}]{Williams2009}
\bibinfo{author}{Williams, D.}, \bibinfo{author}{Carter, B.},
  \bibinfo{year}{2009}.
\newblock \bibinfo{title}{Transmission Electron Microscopy - A Textbook for
  Materials Science}.
\newblock \bibinfo{edition}{2} ed., \bibinfo{publisher}{Springer New York, NY}.
\newblock \DOIprefix\doi{https://doi.org/10.1007/978-0-387-76501-3}.
\bibitem[{Rauch and Véron(2014)}]{Rauch2014}
\bibinfo{author}{Rauch, E.F.}, \bibinfo{author}{Véron, M.},
  \bibinfo{year}{2014}.
\newblock \bibinfo{title}{Automated crystal orientation and phase mapping in
  tem}.
\newblock \bibinfo{journal}{Materials Characterization} \bibinfo{volume}{98},
  \bibinfo{pages}{1--9}.
\newblock \URLprefix
  \url{https://www.sciencedirect.com/science/article/pii/S1044580314002514},
  \DOIprefix\doi{https://doi.org/10.1016/j.matchar.2014.08.010}.
\bibitem[{Wright and Nowell(2005)}]{Wright2005}
\bibinfo{author}{Wright, S.I.}, \bibinfo{author}{Nowell, M.M.},
  \bibinfo{year}{2005}.
\newblock \bibinfo{title}{Ebsd image quality mapping}.
\newblock \bibinfo{journal}{Microscopy and Microanalysis} \bibinfo{volume}{12},
  \bibinfo{pages}{72--84}.
\newblock \URLprefix \url{https://doi.org/10.1017/S1431927606060090},
  \DOIprefix\doi{10.1017/S1431927606060090},
  \href{http://arxiv.org/abs/https://academic.oup.com/mam/article-pdf/12/1/72/48096749/mam0072.pdf}{\tt
  arXiv:https://academic.oup.com/mam/article-pdf/12/1/72/48096749/mam0072.pdf}.
\bibitem[{Schwarzer et~al.(2009)Schwarzer, Field, Adams, Kumar and
  Schwartz}]{Schwarzer2009}
\bibinfo{author}{Schwarzer, R.A.}, \bibinfo{author}{Field, D.P.},
  \bibinfo{author}{Adams, B.L.}, \bibinfo{author}{Kumar, M.},
  \bibinfo{author}{Schwartz, A.J.}, \bibinfo{year}{2009}.
\newblock \bibinfo{title}{Present State of Electron Backscatter Diffraction and
  Prospective Developments}. \bibinfo{publisher}{Springer US},
  \bibinfo{address}{Boston, MA}. chapter~\bibinfo{chapter}{1}.
\newblock pp. \bibinfo{pages}{1--20}.
\newblock \URLprefix \url{https://doi.org/10.1007/978-0-387-88136-2_1},
  \DOIprefix\doi{10.1007/978-0-387-88136-2_1}.
\bibitem[{DeCost et~al.(2020)DeCost, Hattrick-Simpers, Trautt, Kusne, Campo and
  Green}]{DeCost2020}
\bibinfo{author}{DeCost, B.}, \bibinfo{author}{Hattrick-Simpers, J.},
  \bibinfo{author}{Trautt, Z.}, \bibinfo{author}{Kusne, A.},
  \bibinfo{author}{Campo, E.}, \bibinfo{author}{Green, M.},
  \bibinfo{year}{2020}.
\newblock \bibinfo{title}{Scientific ai in materials science: a path to a
  sustainable and scalable paradigm}.
\newblock \bibinfo{journal}{Machine Learning: Science and Technology}
  \bibinfo{volume}{1}, \bibinfo{pages}{033001}.
\newblock \URLprefix \url{https://dx.doi.org/10.1088/2632-2153/ab9a20},
  \DOIprefix\doi{10.1088/2632-2153/ab9a20}.
\bibitem[{Jain(2024)}]{Jain2024}
\bibinfo{author}{Jain, A.}, \bibinfo{year}{2024}.
\newblock \bibinfo{title}{Machine learning in materials research: Developments
  over the last decade and challenges for the future}.
\newblock \bibinfo{journal}{Current Opinion in Solid State and Materials
  Science} \bibinfo{volume}{33}, \bibinfo{pages}{101189}.
\newblock \URLprefix
  \url{https://www.sciencedirect.com/science/article/pii/S135902862400055X},
  \DOIprefix\doi{https://doi.org/10.1016/j.cossms.2024.101189}.
\bibitem[{Kalinin et~al.(2022)Kalinin, Ghosh, Vasudevan and
  Ziatdinov}]{Kalinin2022}
\bibinfo{author}{Kalinin, S.V.}, \bibinfo{author}{Ghosh, A.},
  \bibinfo{author}{Vasudevan, R.}, \bibinfo{author}{Ziatdinov, M.},
  \bibinfo{year}{2022}.
\newblock \bibinfo{title}{From atomically resolved imaging to generative and
  causal models}.
\newblock \bibinfo{journal}{Nature Physics} \bibinfo{volume}{18},
  \bibinfo{pages}{1152--1160}.
\newblock \URLprefix \url{https://doi.org/10.1038/s41567-022-01666-0},
  \DOIprefix\doi{10.1038/s41567-022-01666-0}.
\bibitem[{Ziletti et~al.(2018)Ziletti, Kumar, Scheffler and
  Ghiringhelli}]{Ziletti2018}
\bibinfo{author}{Ziletti, A.}, \bibinfo{author}{Kumar, D.},
  \bibinfo{author}{Scheffler, M.}, \bibinfo{author}{Ghiringhelli, L.M.},
  \bibinfo{year}{2018}.
\newblock \bibinfo{title}{Insightful classification of crystal structures using
  deep learning}.
\newblock \bibinfo{journal}{Nature Communications} \bibinfo{volume}{9},
  \bibinfo{pages}{2775}.
\newblock \URLprefix \url{https://doi.org/10.1038/s41467-018-05169-6},
  \DOIprefix\doi{10.1038/s41467-018-05169-6}.
\bibitem[{Butler et~al.(2018)Butler, Davies, Cartwright, Isayev and
  Walsh}]{Butler2018}
\bibinfo{author}{Butler, K.T.}, \bibinfo{author}{Davies, D.W.},
  \bibinfo{author}{Cartwright, H.}, \bibinfo{author}{Isayev, O.},
  \bibinfo{author}{Walsh, A.}, \bibinfo{year}{2018}.
\newblock \bibinfo{title}{Machine learning for molecular and materials
  science}.
\newblock \bibinfo{journal}{Nature} \bibinfo{volume}{559},
  \bibinfo{pages}{547--555}.
\newblock \URLprefix \url{https://doi.org/10.1038/s41586-018-0337-2},
  \DOIprefix\doi{10.1038/s41586-018-0337-2}.
\bibitem[{Zuo et~al.(2022)Zuo, Yuan, Shao, Hsiao, Pidaparthy, Hu, Yang and
  Zhang}]{Zuo2022}
\bibinfo{author}{Zuo, J.M.}, \bibinfo{author}{Yuan, R.}, \bibinfo{author}{Shao,
  Y.T.}, \bibinfo{author}{Hsiao, H.W.}, \bibinfo{author}{Pidaparthy, S.},
  \bibinfo{author}{Hu, Y.}, \bibinfo{author}{Yang, Q.}, \bibinfo{author}{Zhang,
  J.}, \bibinfo{year}{2022}.
\newblock \bibinfo{title}{Data-driven electron microscopy: electron diffraction
  imaging of materials structural properties}.
\newblock \bibinfo{journal}{Microscopy} \bibinfo{volume}{71},
  \bibinfo{pages}{i116--i131}.
\newblock \URLprefix \url{https://doi.org/10.1093/jmicro/dfab032},
  \DOIprefix\doi{10.1093/jmicro/dfab032},
  \href{http://arxiv.org/abs/https://academic.oup.com/jmicro/article-pdf/71/Supplement\_1/i116/42548319/dfab032.pdf}{\tt
  arXiv:https://academic.oup.com/jmicro/article-pdf/71/Supplement\_1/i116/42548319/dfab032.pdf}.
\bibitem[{Jha et~al.(2018)Jha, Singh, Al-Bahrani, Liao, Choudhary, De~Graef and
  Agrawal}]{Jha2018}
\bibinfo{author}{Jha, D.}, \bibinfo{author}{Singh, S.},
  \bibinfo{author}{Al-Bahrani, R.}, \bibinfo{author}{Liao, W.k.},
  \bibinfo{author}{Choudhary, A.}, \bibinfo{author}{De~Graef, M.},
  \bibinfo{author}{Agrawal, A.}, \bibinfo{year}{2018}.
\newblock \bibinfo{title}{Extracting grain orientations from ebsd patterns of
  polycrystalline materials using convolutional neural networks}.
\newblock \bibinfo{journal}{Microscopy and Microanalysis} \bibinfo{volume}{24},
  \bibinfo{pages}{497--502}.
\newblock \URLprefix \url{https://doi.org/10.1017/S1431927618015131},
  \DOIprefix\doi{10.1017/S1431927618015131},
  \href{http://arxiv.org/abs/https://academic.oup.com/mam/article-pdf/24/5/497/48088526/mam0497.pdf}{\tt
  arXiv:https://academic.oup.com/mam/article-pdf/24/5/497/48088526/mam0497.pdf}.
\bibitem[{Suker(2022)}]{Suker2022}
\bibinfo{author}{Suker, D.K.}, \bibinfo{year}{2022}.
\newblock \bibinfo{title}{Deep learning cnn for the prediction of grain
  orientations on ebsd patterns of aa5083 alloy}.
\newblock \bibinfo{journal}{Engineering, Technology \&amp; Applied Science
  Research} \bibinfo{volume}{12}, \bibinfo{pages}{8393–8401}.
\newblock \URLprefix \url{https://etasr.com/index.php/ETASR/article/view/4807},
  \DOIprefix\doi{10.48084/etasr.4807}.
\bibitem[{Ding et~al.(2020)Ding, Pascal and {De Graef}}]{Ding2020}
\bibinfo{author}{Ding, Z.}, \bibinfo{author}{Pascal, E.}, \bibinfo{author}{{De
  Graef}, M.}, \bibinfo{year}{2020}.
\newblock \bibinfo{title}{Indexing of electron back-scatter diffraction
  patterns using a convolutional neural network}.
\newblock \bibinfo{journal}{Acta Materialia} \bibinfo{volume}{199},
  \bibinfo{pages}{370--382}.
\newblock \URLprefix
  \url{https://www.sciencedirect.com/science/article/pii/S1359645420306510},
  \DOIprefix\doi{https://doi.org/10.1016/j.actamat.2020.08.046}.
\bibitem[{Hara et~al.(2023)Hara, Kojima, Kutsukake, Kudo and Usami}]{Hara2023}
\bibinfo{author}{Hara, K.}, \bibinfo{author}{Kojima, T.},
  \bibinfo{author}{Kutsukake, K.}, \bibinfo{author}{Kudo, H.},
  \bibinfo{author}{Usami, N.}, \bibinfo{year}{2023}.
\newblock \bibinfo{title}{A machine learning-based prediction of crystal
  orientations for multicrystalline materials}.
\newblock \bibinfo{journal}{APL Machine Learning} \bibinfo{volume}{1},
  \bibinfo{pages}{026113}.
\newblock \URLprefix \url{https://doi.org/10.1063/5.0138099},
  \DOIprefix\doi{10.1063/5.0138099},
  \href{http://arxiv.org/abs/https://pubs.aip.org/aip/aml/article-pdf/doi/10.1063/5.0138099/17776291/026113\_1\_5.0138099.pdf}{\tt
  arXiv:https://pubs.aip.org/aip/aml/article-pdf/doi/10.1063/5.0138099/17776291/026113\_1\_5.0138099.pdf}.
\bibitem[{Shen et~al.(2019)Shen, Pokharel, Nizolek, Kumar and
  Lookman}]{Shen2019}
\bibinfo{author}{Shen, Y.F.}, \bibinfo{author}{Pokharel, R.},
  \bibinfo{author}{Nizolek, T.J.}, \bibinfo{author}{Kumar, A.},
  \bibinfo{author}{Lookman, T.}, \bibinfo{year}{2019}.
\newblock \bibinfo{title}{Convolutional neural network-based method for
  real-time orientation indexing of measured electron backscatter diffraction
  patterns}.
\newblock \bibinfo{journal}{Acta Materialia} \bibinfo{volume}{170},
  \bibinfo{pages}{118--131}.
\newblock \URLprefix
  \url{https://www.sciencedirect.com/science/article/pii/S1359645419301697},
  \DOIprefix\doi{https://doi.org/10.1016/j.actamat.2019.03.026}.
\bibitem[{Yuan et~al.(2021)Yuan, Zhang, He and Zuo}]{Yuan2021}
\bibinfo{author}{Yuan, R.}, \bibinfo{author}{Zhang, J.}, \bibinfo{author}{He,
  L.}, \bibinfo{author}{Zuo, J.M.}, \bibinfo{year}{2021}.
\newblock \bibinfo{title}{Training artificial neural networks for precision
  orientation and strain mapping using 4d electron diffraction datasets}.
\newblock \bibinfo{journal}{Ultramicroscopy} \bibinfo{volume}{231},
  \bibinfo{pages}{113256}.
\newblock \URLprefix
  \url{https://www.sciencedirect.com/science/article/pii/S0304399121000486},
  \DOIprefix\doi{https://doi.org/10.1016/j.ultramic.2021.113256}.
  \bibinfo{note}{80th Birthdays of Colin Humphreys and Knut Urban, 75th
  Birthdays of Wolfgang Baumeister and John Spence - PICO 2021 – Sixth
  Conference on Frontiers of Aberration Corrected Electron Microscopy}.
\bibitem[{Demuth et~al.(2025)Demuth, Kurzhals, Ahmed, Riewald, Malaki, Haust,
  Beyer, Janek and Volz}]{Demuth2025}
\bibinfo{author}{Demuth, T.}, \bibinfo{author}{Kurzhals, P.},
  \bibinfo{author}{Ahmed, S.}, \bibinfo{author}{Riewald, F.},
  \bibinfo{author}{Malaki, M.}, \bibinfo{author}{Haust, J.},
  \bibinfo{author}{Beyer, A.}, \bibinfo{author}{Janek, J.},
  \bibinfo{author}{Volz, K.}, \bibinfo{year}{2025}.
\newblock \bibinfo{title}{Effect of a two-step temperature-swing synthesis on
  coarse-grained linio2 secondary particles characterized by scanning
  transmission electron microscopy}.
\newblock \bibinfo{journal}{Chemistry of Materials} \bibinfo{volume}{37},
  \bibinfo{pages}{3993--4004}.
\newblock \URLprefix \url{https://doi.org/10.1021/acs.chemmater.5c00108},
  \DOIprefix\doi{10.1021/acs.chemmater.5c00108},
  \href{http://arxiv.org/abs/https://doi.org/10.1021/acs.chemmater.5c00108}{\tt
  arXiv:https://doi.org/10.1021/acs.chemmater.5c00108}.
\bibitem[{Rauch et~al.(2010)Rauch, Portillo, Nicolopoulos, Bultreys, Rouvimov
  and Moeck}]{Rauch2010}
\bibinfo{author}{Rauch, E.F.}, \bibinfo{author}{Portillo, J.},
  \bibinfo{author}{Nicolopoulos, S.}, \bibinfo{author}{Bultreys, D.},
  \bibinfo{author}{Rouvimov, S.}, \bibinfo{author}{Moeck, P.},
  \bibinfo{year}{2010}.
\newblock \bibinfo{title}{Automated nanocrystal orientation and phase mapping
  in the transmission electron microscope on the basis of precession electron
  diffraction}.
\newblock \bibinfo{journal}{Zeitschrift für Kristallographie}
  \bibinfo{volume}{225}, \bibinfo{pages}{103--109}.
\newblock \URLprefix \url{https://doi.org/10.1524/zkri.2010.1205},
  \DOIprefix\doi{doi:10.1524/zkri.2010.1205}.
\bibitem[{Rauch et~al.(2021)Rauch, Harrison, Zhou, Herbig, Ludwig and
  Véron}]{Rauch2021}
\bibinfo{author}{Rauch, E.F.}, \bibinfo{author}{Harrison, P.},
  \bibinfo{author}{Zhou, X.}, \bibinfo{author}{Herbig, M.},
  \bibinfo{author}{Ludwig, W.}, \bibinfo{author}{Véron, M.},
  \bibinfo{year}{2021}.
\newblock \bibinfo{title}{New features in crystal orientation and phase mapping
  for transmission electron microscopy}.
\newblock \bibinfo{journal}{Symmetry} \bibinfo{volume}{13}.
\newblock \URLprefix \url{https://www.mdpi.com/2073-8994/13/9/1675},
  \DOIprefix\doi{10.3390/sym13091675}.
\bibitem[{Scheunert et~al.(2025)Scheunert, Ahmed, Demuth, Beyer, Wissel, Xu and
  Volz}]{Scheunert2025}
\bibinfo{author}{Scheunert, J.}, \bibinfo{author}{Ahmed, S.},
  \bibinfo{author}{Demuth, T.}, \bibinfo{author}{Beyer, A.},
  \bibinfo{author}{Wissel, S.}, \bibinfo{author}{Xu, B.X.},
  \bibinfo{author}{Volz, K.}, \bibinfo{year}{2025}.
\newblock \bibinfo{title}{Determining the grain orientations of battery
  materials from electron diffraction patterns using convolutional neural
  networks}.
\newblock \URLprefix \url{https://arxiv.org/abs/2506.18416},
  \href{http://arxiv.org/abs/2506.18416}{\tt arXiv:2506.18416}.
\bibitem[{Johnstone et~al.(2020)Johnstone, Martineau, Crout, Midgley and
  Eggeman}]{Johnstone2020}
\bibinfo{author}{Johnstone, D.N.}, \bibinfo{author}{Martineau, B.H.},
  \bibinfo{author}{Crout, P.}, \bibinfo{author}{Midgley, P.A.},
  \bibinfo{author}{Eggeman, A.S.}, \bibinfo{year}{2020}.
\newblock \bibinfo{title}{{Density-based clustering of crystal
  (mis)orientations and the {\it orix} Python library}}.
\newblock \bibinfo{journal}{Journal of Applied Crystallography}
  \bibinfo{volume}{53}, \bibinfo{pages}{1293--1298}.
\newblock \URLprefix \url{https://doi.org/10.1107/S1600576720011103},
  \DOIprefix\doi{10.1107/S1600576720011103}.
\bibitem[{Ånes et~al.(2025)Ånes, Martineau, Harrison, Crout, Johnstone,
  Cautaerts, Gerlt, Mathisen, Xu, Francis, Høgås, Femoen, da~Silva and
  Clausen}]{Aanes2025}
\bibinfo{author}{Ånes, H.W.}, \bibinfo{author}{Martineau, B.},
  \bibinfo{author}{Harrison, P.}, \bibinfo{author}{Crout, P.},
  \bibinfo{author}{Johnstone, D.}, \bibinfo{author}{Cautaerts, N.},
  \bibinfo{author}{Gerlt, A.}, \bibinfo{author}{Mathisen, A.C.},
  \bibinfo{author}{Xu, Z.}, \bibinfo{author}{Francis, C.},
  \bibinfo{author}{Høgås, S.}, \bibinfo{author}{Femoen, V.J.},
  \bibinfo{author}{da~Silva, A.}, \bibinfo{author}{Clausen, A.},
  \bibinfo{year}{2025}.
\newblock \bibinfo{title}{pyxem/orix: orix 0.13.3}.
\newblock \DOIprefix\doi{10.5281/ZENODO.14601499}.
\bibitem[{Huber(1964)}]{Huber1964}
\bibinfo{author}{Huber, P.J.}, \bibinfo{year}{1964}.
\newblock \bibinfo{title}{{Robust Estimation of a Location Parameter}}.
\newblock \bibinfo{journal}{The Annals of Mathematical Statistics}
  \bibinfo{volume}{35}, \bibinfo{pages}{73 -- 101}.
\newblock \URLprefix \url{https://doi.org/10.1214/aoms/1177703732},
  \DOIprefix\doi{10.1214/aoms/1177703732}.
\bibitem[{Kingma and Ba(2017)}]{Kingma2017}
\bibinfo{author}{Kingma, D.P.}, \bibinfo{author}{Ba, J.}, \bibinfo{year}{2017}.
\newblock \bibinfo{title}{Adam: A method for stochastic optimization}.
\newblock \URLprefix \url{https://arxiv.org/abs/1412.6980},
  \href{http://arxiv.org/abs/1412.6980}{\tt arXiv:1412.6980}.
\bibitem[{Indolia et~al.(2018)Indolia, Goswami, Mishra and Asopa}]{Indolia2018}
\bibinfo{author}{Indolia, S.}, \bibinfo{author}{Goswami, A.K.},
  \bibinfo{author}{Mishra, S.}, \bibinfo{author}{Asopa, P.},
  \bibinfo{year}{2018}.
\newblock \bibinfo{title}{Conceptual understanding of convolutional neural
  network- a deep learning approach}.
\newblock \bibinfo{journal}{Procedia Computer Science} \bibinfo{volume}{132},
  \bibinfo{pages}{679--688}.
\newblock \URLprefix
  \url{https://www.sciencedirect.com/science/article/pii/S1877050918308019},
  \DOIprefix\doi{https://doi.org/10.1016/j.procs.2018.05.069}.
  \bibinfo{note}{international Conference on Computational Intelligence and
  Data Science}.
\bibitem[{O'Shea and Nash(2015)}]{Oshea2015}
\bibinfo{author}{O'Shea, K.}, \bibinfo{author}{Nash, R.}, \bibinfo{year}{2015}.
\newblock \bibinfo{title}{An introduction to convolutional neural networks}.
\newblock \URLprefix \url{https://arxiv.org/abs/1511.08458},
  \href{http://arxiv.org/abs/1511.08458}{\tt arXiv:1511.08458}.
\bibitem[{Li et~al.(2022)Li, Liu, Yang, Peng and Zhou}]{Li2022}
\bibinfo{author}{Li, Z.}, \bibinfo{author}{Liu, F.}, \bibinfo{author}{Yang,
  W.}, \bibinfo{author}{Peng, S.}, \bibinfo{author}{Zhou, J.},
  \bibinfo{year}{2022}.
\newblock \bibinfo{title}{A survey of convolutional neural networks: Analysis,
  applications, and prospects}.
\newblock \bibinfo{journal}{IEEE Transactions on Neural Networks and Learning
  Systems} \bibinfo{volume}{33}, \bibinfo{pages}{6999--7019}.
\newblock \DOIprefix\doi{10.1109/TNNLS.2021.3084827}.
\bibitem[{Huang et~al.(2018)Huang, Liu, van~der Maaten and
  Weinberger}]{Huang2018}
\bibinfo{author}{Huang, G.}, \bibinfo{author}{Liu, Z.},
  \bibinfo{author}{van~der Maaten, L.}, \bibinfo{author}{Weinberger, K.Q.},
  \bibinfo{year}{2018}.
\newblock \bibinfo{title}{Densely connected convolutional networks}.
\newblock \URLprefix \url{https://arxiv.org/abs/1608.06993},
  \href{http://arxiv.org/abs/1608.06993}{\tt arXiv:1608.06993}.
\bibitem[{Liu et~al.(2021)Liu, Lin, Cao, Hu, Wei, Zhang, Lin and Guo}]{Liu2021}
\bibinfo{author}{Liu, Z.}, \bibinfo{author}{Lin, Y.}, \bibinfo{author}{Cao,
  Y.}, \bibinfo{author}{Hu, H.}, \bibinfo{author}{Wei, Y.},
  \bibinfo{author}{Zhang, Z.}, \bibinfo{author}{Lin, S.}, \bibinfo{author}{Guo,
  B.}, \bibinfo{year}{2021}.
\newblock \bibinfo{title}{Swin transformer: Hierarchical vision transformer
  using shifted windows}.
\newblock \URLprefix \url{https://arxiv.org/abs/2103.14030},
  \href{http://arxiv.org/abs/2103.14030}{\tt arXiv:2103.14030}.
\bibitem[{Liu et~al.(2022)Liu, Hu, Lin, Yao, Xie, Wei, Ning, Cao, Zhang, Dong,
  Wei and Guo}]{Liu2022}
\bibinfo{author}{Liu, Z.}, \bibinfo{author}{Hu, H.}, \bibinfo{author}{Lin, Y.},
  \bibinfo{author}{Yao, Z.}, \bibinfo{author}{Xie, Z.}, \bibinfo{author}{Wei,
  Y.}, \bibinfo{author}{Ning, J.}, \bibinfo{author}{Cao, Y.},
  \bibinfo{author}{Zhang, Z.}, \bibinfo{author}{Dong, L.},
  \bibinfo{author}{Wei, F.}, \bibinfo{author}{Guo, B.}, \bibinfo{year}{2022}.
\newblock \bibinfo{title}{Swin transformer v2: Scaling up capacity and
  resolution}.
\newblock \URLprefix \url{https://arxiv.org/abs/2111.09883},
  \href{http://arxiv.org/abs/2111.09883}{\tt arXiv:2111.09883}.
\bibitem[{Ding et~al.(2022)Ding, Zhang, Zhou, Han, Ding and Sun}]{Ding2022}
\bibinfo{author}{Ding, X.}, \bibinfo{author}{Zhang, X.}, \bibinfo{author}{Zhou,
  Y.}, \bibinfo{author}{Han, J.}, \bibinfo{author}{Ding, G.},
  \bibinfo{author}{Sun, J.}, \bibinfo{year}{2022}.
\newblock \bibinfo{title}{Scaling up your kernels to 31x31: Revisiting large
  kernel design in cnns}.
\newblock \URLprefix \url{https://arxiv.org/abs/2203.06717},
  \href{http://arxiv.org/abs/2203.06717}{\tt arXiv:2203.06717}.
\end{thebibliography}

\newpage
\appendix
\section*{Appendix}
\renewcommand{\thefigure}{A.\arabic{figure}}
\setcounter{figure}{0}
\renewcommand{\thetable}{A.\arabic{table}}
\setcounter{table}{0}

\section{Virtual Darkfield Image and TM data}\label{a:VDF_TMQ}
\begin{figure}[h]
    \centering
    \includegraphics[width=0.75\textwidth]{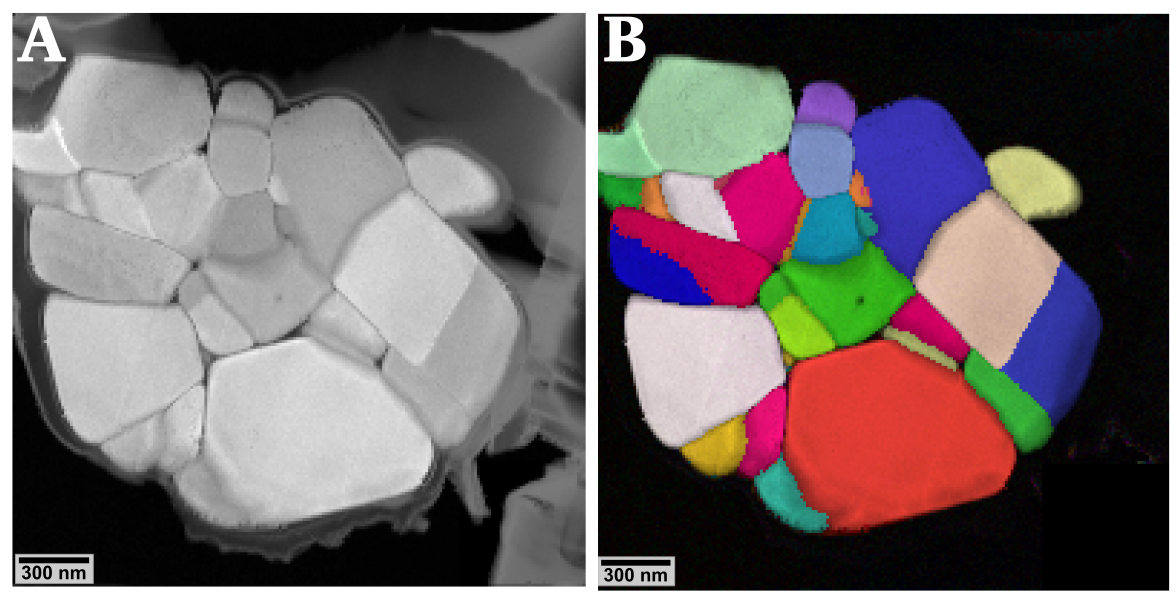}
    \caption{Virtual darkfield image of the experimental dataset (A) and the assigned TM labels weighted with the corresponding $Q$-value (B).}
    \label{fig:VDF_TMQ}
\end{figure}

To provide a more intuitive visual representation of the investigated sample, a virtual darkfield image is generated from the recorded DPs, as shown in Fig.~\ref{fig:VDF_TMQ}~A. In addition, Fig.~\ref{fig:VDF_TMQ}~B presents the assigned TM labels, where each label is weighted by its corresponding correlation index $Q$. In this visualization, higher intensity indicates a higher $Q$ value, reflecting greater confidence in the label assignment.

\FloatBarrier

\section{Evaluation Scores}\label{a:eval}
To quantify model performance, three standard regression metrics were used: \textit{mean absolute error} (MAE), \textit{mean squared error} (MSE), and the \textit{coefficient of determination} ($R^2$-score). These metrics assess different aspects of predictive quality—MAE reflects the average magnitude of errors, MSE penalizes larger deviations more strongly, and the $R^2$-score provides a measure of the proportion of variance explained by the model. The mathematical definitions are given below:
\begin{align*}
    MAE &= \frac{1}{n}\sum_{i=1}^{n}|e_{i}| \\
    MSE &= \frac{1}{n}\sum_{i=1}^{n} e_{i}^{2} \\
    R^{2} &= 1 - \sum_{i=1}^{n} \frac{e_{i}^{2}}{(T_{i}-\overline{T})^{2}}
\end{align*}

where $n$ is the number of samples, $T_i$ the true target, $\overline{T}$ the mean of all targets, $P_i$ the prediction, and $e_i = T_i - P_i$ the error.
\FloatBarrier

\section{Comparison of DenseNets and SwinT Model Predictions}\label{a:comp_DN_SwinT}
In addition to quantitative performance metrics, a qualitative evaluation of the predicted crystal maps was conducted to assess physical plausibility. Specifically, predictions from the two DenseNets models (plain and pre-trained) and four SwinT variants were compared visually for dataset $DS_Q$.

\begin{figure}
    \centering
    \includegraphics[width=0.75\textwidth]{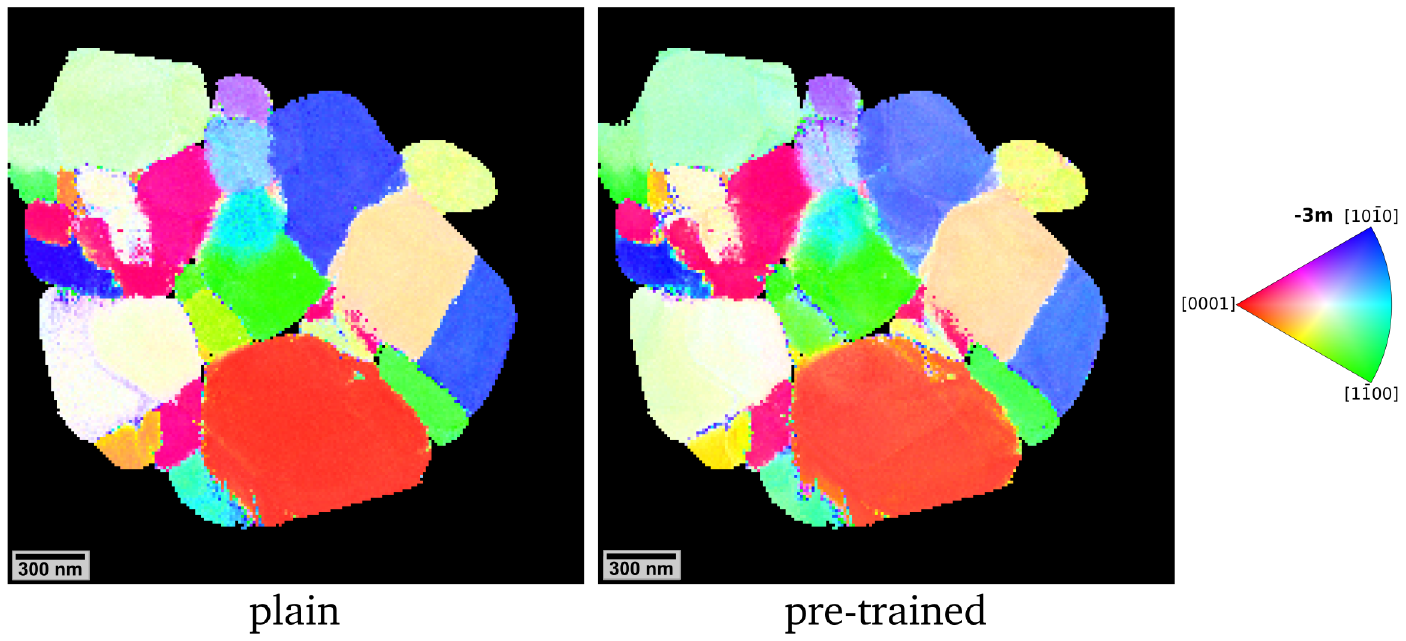}
    \caption{Comparison of $DS_Q$ predictions from the two trained DenseNets models.}
    \label{fig:comp_DN}
\end{figure}

As summarized in Table~\ref{tab:eval_scores}, all models achieved comparable $R^2$ scores across validation, test, and full prediction datasets. However, visual inspection of the predicted orientation maps revealed meaningful differences. The \textit{plain} DenseNets model produced sharper grain boundaries and more homogeneous predictions within individual grains compared to its pre-trained counterpart (Fig.~\ref{fig:comp_DN}). 

\begin{figure}
    \centering
    \includegraphics[width=0.75\textwidth]{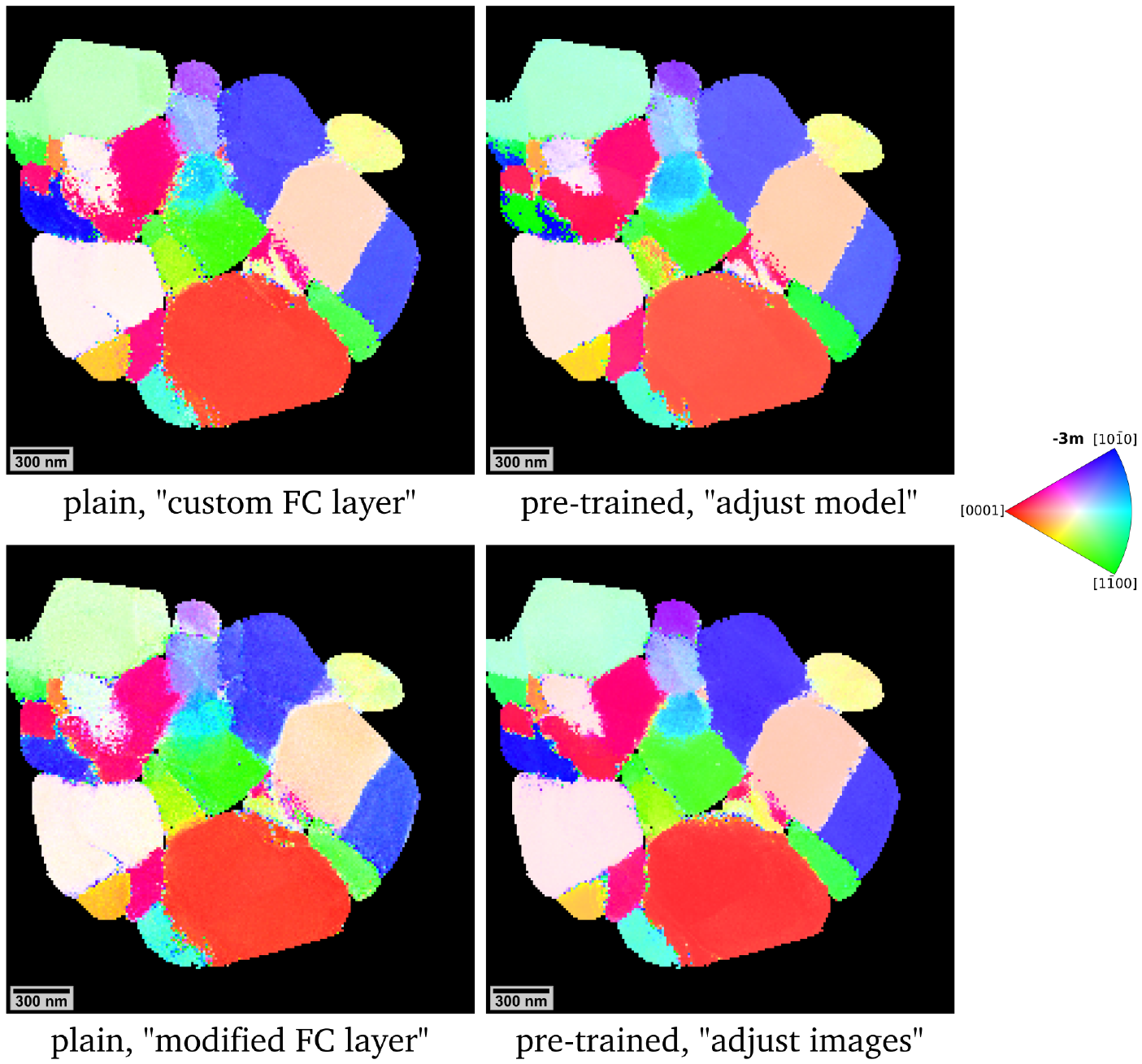}
    \caption{Comparison of $DS_Q$ predictions from the four SwinT model variants.}
    \label{fig:comp_SwinTs}
\end{figure}

A similar pattern was observed among the SwinT models. Despite comparable quantitative scores, the pre-trained SwinT model with an adjusted architecture generated the most coherent orientation maps, exhibiting well-defined grain boundaries and consistent orientation within grains (Fig.~\ref{fig:comp_SwinTs}). These observations emphasize the value of combining metric-based evaluation with visual and physical interpretability of the predicted microstructures.

\begin{table}
\centering
\caption{Evaluation scores of DenseNets and SwinT models across different datasets.}
\label{tab:eval_scores}
\setlength{\tabcolsep}{4pt}
\renewcommand{\arraystretch}{1.1}
\begin{tabular}{lccc}
\toprule
Model & $R^2_\text{Val}$ & $R^2_\text{Test}$ & $R^2_{DS_Q}$ \\
\midrule
DenseNet, plain & 0.96 & 0.95 & 0.92 \\
DenseNet, pre-trained & 0.96 & 0.95 & 0.91 \\
\midrule
SwinT, plain (custom FC layer) & 0.95 & 0.94 & 0.91 \\
SwinT, plain (modified FC layer) & 0.95 & 0.94 & 0.91 \\
SwinT, pre-trained (adjusted images) & 0.95 & 0.95 & 0.91 \\
SwinT, pre-trained (adjusted model) & 0.96 & 0.95 & 0.92 \\
\bottomrule
\end{tabular}
\end{table}

\FloatBarrier

\section{Inverse Pole Figures}\label{a:IPFs}
\begin{figure}
    \centering
    \includegraphics[width=0.75\textwidth]{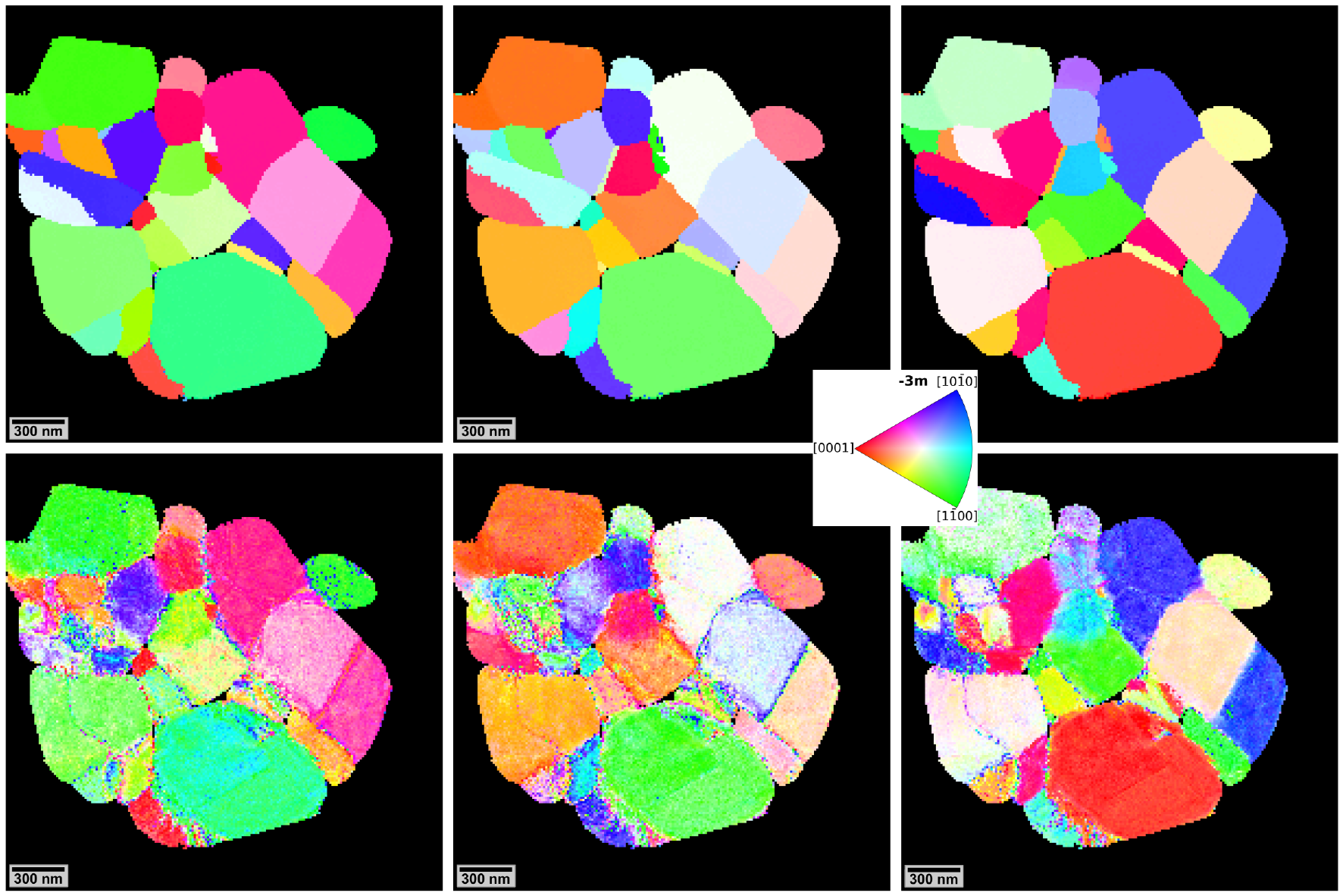}
    \caption{IPFs from TM labels and CNN predictions. The TM labels (top) and predictions of the CNN model (bottom) are shown as IPFs for all three spatial directions.}
    \label{fig:TM_CNN}
\end{figure}

\begin{figure}
    \centering
    \includegraphics[width=0.75\textwidth]{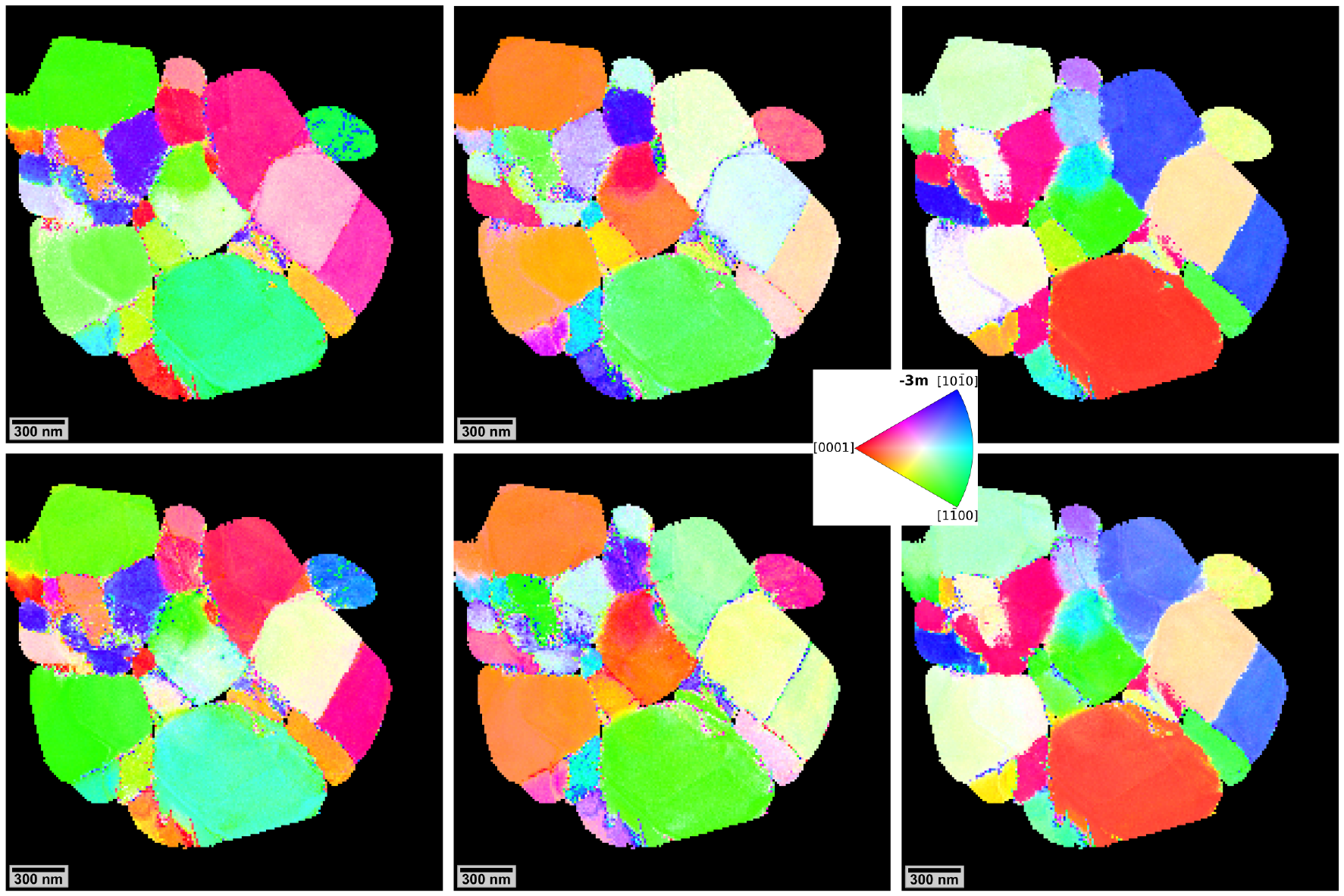}
    \caption{IPFs of the DenseNets model predictions. The DenseNets model predictions of the plain (top) and pre-trained (bottom) model are shown as IPFs for all three spatial directions.}
    \label{fig:DenseNet}
\end{figure}

\begin{figure}
    \centering
    \includegraphics[width=0.75\textwidth]{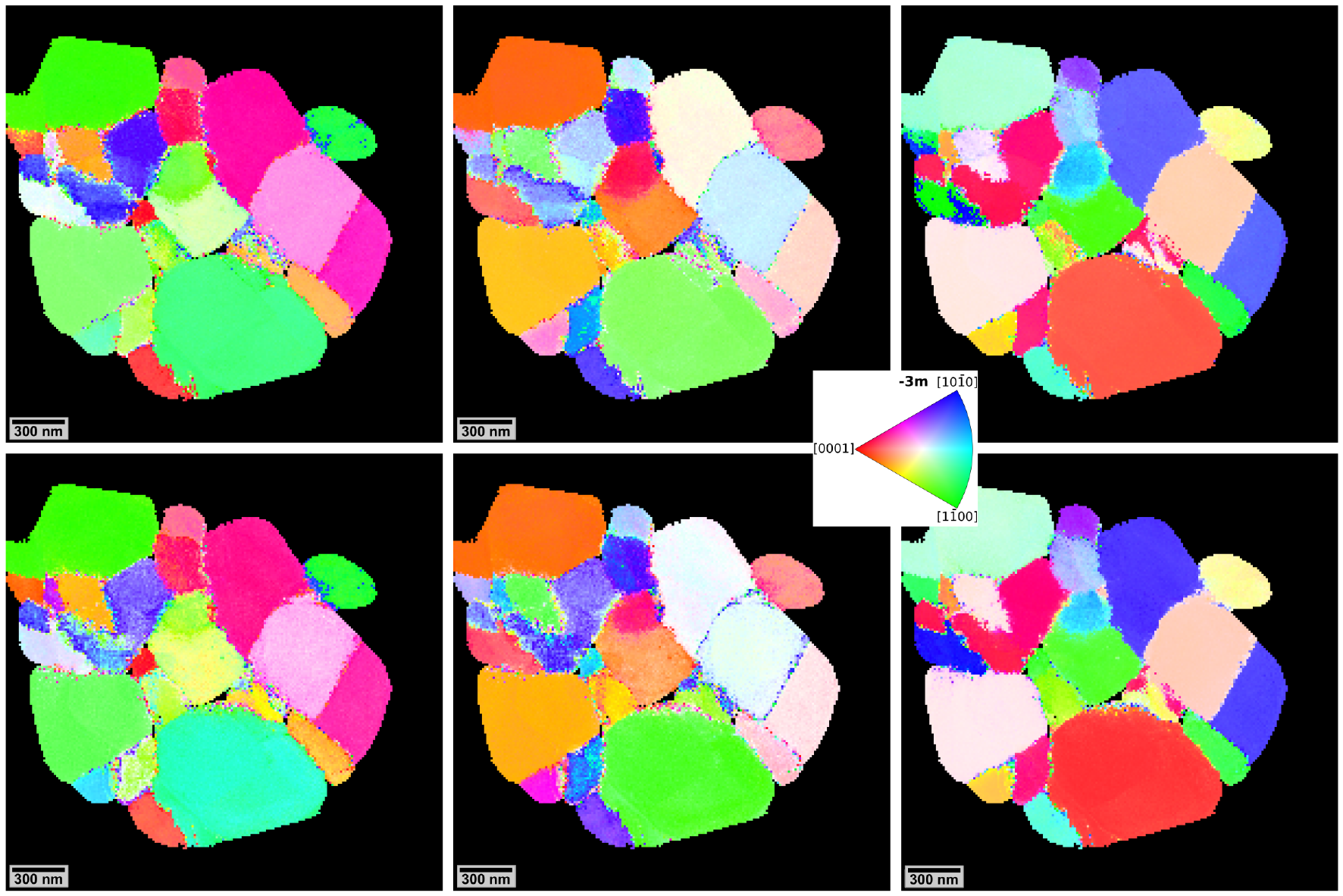}
    \caption{IPFs of the pre-trained SwinT model predictions. The predictions of the pre-trained SwinT models "adjust model" (top) and "adjust images" (bottom) are shown as IPFs for all three spatial directions.}
    \label{fig:SwinT_pretrained}
\end{figure}

\begin{figure}
    \centering
    \includegraphics[width=0.75\textwidth]{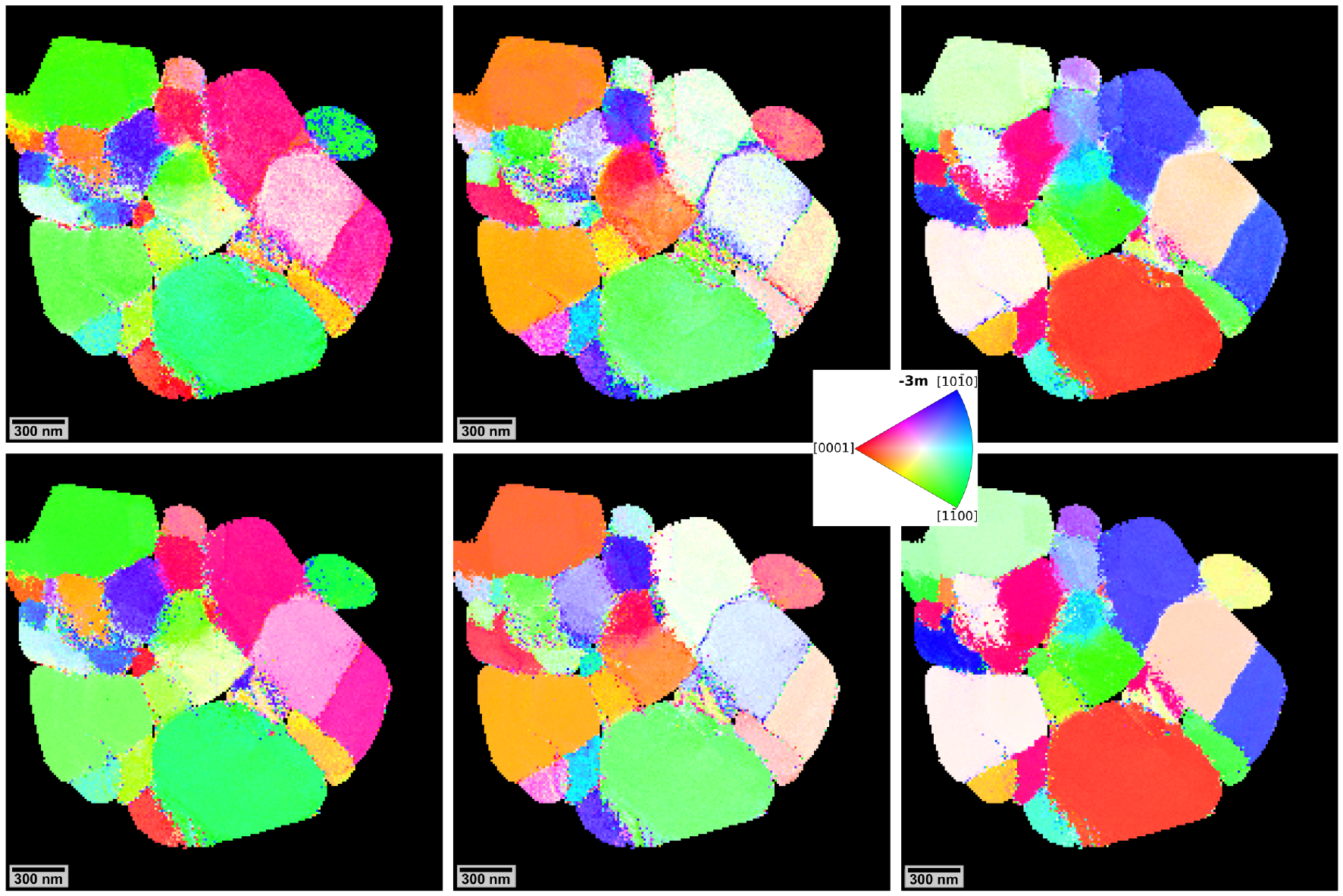}
    \caption{IPFs of the plain SwinT model predictions. The predictions of the plain SwinT models "modified FC layer" (top) and "custom FC layer" (bottom) are shown as IPFs for all three spatial directions.}
    \label{fig:SwinT_plain}
\end{figure}

To fully characterize the crystallographic orientation of grains, IPFs must be evaluated with respect to all three spatial axes: $x$, $y$, and $z$. While the main text focuses on $z$-axis IPFs, the $x$- and $y$-axis projections provide essential complementary information. The figures \ref{fig:TM_CNN} (TM labels and CNN predictions), \ref{fig:DenseNet} (DenseNets predictions), \ref{fig:SwinT_pretrained} and \ref{fig:SwinT_plain} (SwinT predictions) present the full sets of IPFs for each model variant. 

\end{document}